\documentclass[fleqn,usenatbib]{mnras}

\usepackage{newtxtext,newtxmath}
\usepackage[T1]{fontenc}
\usepackage{ae,aecompl}

\usepackage{graphicx}	% Including figure files
\usepackage{amsmath}	% Advanced maths commands
\usepackage{physics}
\usepackage{float}
\usepackage{longtable}
\usepackage{soul}

\newcommand{\parx}{K_\mathrm{dK}}
\newcommand{\pary}{v_\gamma}
\newcommand{\parz}{\phi_0}
\newcommand{\dO}{\mathrm{d}\Omega}
\newcommand{\Oeq}{\Omega_\mathrm{eq}}
\newcommand{\rchi}{\chi^2}

\title[Magnetic field of the K2 dwarf V471~Tau]{Magnetic field $\&$ activity phenomena of the K2 dwarf V471 Tau}

\author[B.~Zaire et. al]{
B.~Zaire\thanks{E-mail: bonnie.zaire@irap.omp.eu},
J.-F.~Donati,
and  B.~Klein
\\
IRAP, Université de Toulouse, CNRS / UMR 5277, CNES, UPS, 14 avenue E. Belin, Toulouse, F-31400 France}

\date{Accepted XXX. Received YYY; in original form ZZZ}

\pubyear{2020}

\begin{document}
\label{firstpage}
\pagerange{\pageref{firstpage}--\pageref{lastpage}}
\maketitle

\begin{abstract}
We analyze spectropolarimetric data of the pre-cataclysmic variable binary system V471~Tau obtained with ESPaDOnS at the Canada-France-Hawaii Telescope in two observational campaigns (in Nov/Dec 2004 and Dec 2005). Using Zeeman-Doppler Imaging, we reconstruct the distribution of brightness map and large-scale magnetic field of the K2 dwarf at both epochs, as well as the amount of differential rotation by which surface maps are sheared. We detect significant fluctuations in the surface shear between the two campaigns. It goes from about twice the solar differential rotation rate to less than the solar value in a one-year interval. We conclude that the differential rotation fluctuations obtained for the K2 dwarf resemble those detected on the single-star analog AB~Dor,  although even larger amplitudes of variation are seen in the K2 dwarf of V471~Tau. Finally, we show that the differential rotation results obtained in this work do not favor an Applegate mechanism operating in the V471 Tau system, at least in its standard form, but leave room for explaining the observed orbital period fluctuations with exotic forms of similar phenomena based on dynamo processes operating within the convective zone of the K2 star.
\end{abstract}


%%%%%%%%%%%%%%%%%%%%%%%%%%%%%%%%%%%%%%%%%%%%%%%%%%

%%%%%%%%%%%%%%%%% BODY OF PAPER %%%%%%%%%%%%%%%%%%

%%%%%%%%%%%%%%%%%%%%%%%%%%%%%%%%%%%%%%%%%%%
\section{Introduction}
%%%%%%%%%%%%%%%%%%%%%%%%%%%%%%%%%%%%%%%%%%%
Over the past decade, magnetic fields have been studied for stars in a wide range of spectral classes, with a specific focus on solar-like stars of various masses, rotation rates, and ages. It has conclusively been shown that the field strength decays with age for solar-type stars \citep{VGJ14}, obeying a similar power-law to what has been obtained for the rotational period \citep{S72}. This dependence between the magnetic properties and evolutionary state implies that the dynamo mechanisms that are at play in the convective zone also evolve and adjust throughout the star's life.

Thanks to its ability to reconstruct large-scale surface magnetic fields, Zeeman-Doppler Imaging technique \citep[hereafter ZDI,][]{DB97} brought new insights to the subject establishing the critical role of the stellar internal structure on the overall topology of the large-scale field \citep{GDM12}. Typically, fully-convective stars were found to harbor strong poloidal magnetic fields with a high degree of axisymmetry, while partly-convective stars displayed weaker surface fields with a predominantly non-axisymmetric poloidal component and a significant (sometimes dominant) toroidal component \citep[e.g., ][]{DMP08, MDP08, MDP10,FPB16,FBP18}. This scenario gives hints on how dynamo processes may be acting for a range of stellar parameters - with mostly single stars studied so far. 

AB~Dor is the first star on which differential rotation was detected, and it has been the subject of a large number of dedicated observing campaigns due to its key role in addressing the dynamo action in stars \citep[e.g.,][]{DCP03,JDC07}. The K2 dwarf of the V471~Tau binary system is a twin version of the single-star AB~Dor. As a member of a close binary system, involving a white dwarf of mass similar to the K2 dwarf \citep{VWV15}, tidal forces are expected to directly impact the angular momentum evolution, aiming to synchronize the rotational period with the orbital period of the system \citep{Z89}. It has been argued that tidal effects would weaken the differential rotation of the active star \citep{S81,S82}; however, the strong shear detected in the pre-main-sequence binary system HD~155555 opposes this claim \citep{DHC08}. It is still unknown whether the young age of the HD~155555 system (18~Myr) could explain the observed shear. Nevertheless, observations suggest the existence of a preferential longitude for the spots manifestation in short-period binaries, likely due to the influence of the tides on the dynamo action \citep[see][for a discussion on the formation of preferential longitudes]{HS02,HS03,HS03B}. 
 
Being also an eclipsing binary, V471 Tau allows studying the potential impact of dynamo action and activity cycles on the observed eclipse timing variations (ETV). Although the ETVs of compact binaries are commonly associated with a third body, e.g., for NN Ser \citep{MPB14,BDH13} and QS Vir \citep{PMC10,BMP16}, in V471~Tau some authors questioned the presence of the brown dwarf necessary to explain the ETVs \citep{HSP15,VCD17}. As first proposed by \citet{A92}, stars exhibiting magnetic cycles might be capable of changing the internal mass distribution (or equivalently the quadrupole moment of the star) by redistributing angular momentum in the convective envelope. This effect is of particular interest for close binary systems, where the resulting gravity change of a companion propagates to the system, culminating in orbital period modulations. However, improvements in Applegate models challenged its feasibility to drive ETV \citep{L05,L06,VSB18}. Specifically, the inclusion of more realistic angular velocities (with radial and latitudinal dependencies) limited the Applegate effect to systems with large shear fluctuations, thus making it unlikely for most binary systems. 

An alternative mechanism requiring low levels of fluctuation in the differential rotation was recently suggested to operate in V471~Tau \citep{L20}. The mechanism relies on the existence of a stationary non-axisymmetric quadrupole moment in the K2 dwarf that either librates or circulates in the orbital plane. Contrary to the Applegate effect, \citeauthor{L20}'s model requires a stationary non-axisymmetric magnetic field to sustain the quadrupole moment and drive ETV (a result of a torque introduced by magnetic structures misaligned with the line joining both companions). For V471~Tau, \citet{L20} showed that both libration and circulation scenarios could be operating in the system.
 
In the present work, we investigate the magnetism of the cool companion of the close-binary system V471~Tau by analyzing spectropolarimetric observations collected with ESPaDOnS in Nov/Dec 2004 and Dec 2005. This study offers a unique opportunity to examine how tides affect the magnetic topology and differential rotation compared to the single-star analog AB~Dor (similar mass, temperature, and rotational period). The evolutionary status of the system is discussed in Sec.~\ref{sec:paramsv471tau} and our data set is presented in Sec.~\ref{sec:obsdata}. After describing the ZDI technique in Sec.~\ref{sec:modelling}, we present the reconstructed maps of brightness inhomogeneities and large-scale magnetic field in Sec.~\ref{sec:maps}. Section~\ref{sec:diff} is dedicated to the positive detection of differential rotation at the surface of the star, and Sec.~\ref{sec:halpha} to the periodic behavior observed in the H$\alpha$ emission line for both epochs of observation. Finally, we discuss our results in Sec.~\ref{sec:disc}.

%%%%%%%%%%%%%%%%%%%%%%%%%%%%%%%%%%%%%%%%%%%
\section{Evolutionary Stage of V471 Tau} \label{sec:paramsv471tau}
%%%%%%%%%%%%%%%%%%%%%%%%%%%%%%%%%%%%%%%%%%%

V471~Tau is an eclipsing binary system and member of the 625~Myr old Hyades open cluster \citep{PBL98}  with a \textit{Gaia} distance of $47.51 \pm 0.03$~pc \citep{G20,BRF21}. Over the last $50$~years, V471~Tau has been extensively observed to understand the evolution of binary systems with the eclipses giving a unique opportunity to measure the orbital period of the system and its temporal variation with extreme accuracy. The current scenario indicates that the system is a pre-cataclysmic variable that has undergone a common-envelope phase in the early stages of evolution. The system consists of a hot white dwarf star (WD) and a K2 dwarf main-sequence star not yet overfilling its Roche lobe \citep{NY70}. Self-consistent analysis handling simultaneously radial velocity curves, light curves, and eclipse timings of the system yielded a WD mass of $0.8778 \pm 0.0011$~M$_{\sun}$ and a K2 dwarf mass of $0.9971 \pm 0.0012$~M$_{\sun}$, orbiting with a short-period of $P_\mathrm{orb} = 0.5211833875$~days and a separation distance of $a = 3.586~R_\star$, where $R_\star$ is the radius of the K2 dwarf \citep{VWV15}. Moreover, because of the proximity of the two companions, tides compel the K2 dwarf star to rotate synchronously with the orbital period of the system, implying that $P_\mathrm{rot} \simeq P_\mathrm{orb}$. We summarise the quantities relevant for the scope of this paper in Table~\ref{tab:paramsv471}.

\begin{table}
\addtolength{\tabcolsep}{-2pt}
    \centering
    \caption{Parameters of the K2 dwarf component of the V471~Tau system. From top to bottom: age, distance from the Earth $d$, separation distance to the companion $a$, mass $M_\star$, radius $R_\star$, effective temperature ${T_\mathrm{eff}}$, logarithm of the surface gravity $\log g$, rotational period $P_\mathrm{rot}$, inclination $i$, and line-of-sight projected equatorial rotation velocity $v\sin(i)$. }
    \label{tab:paramsv471}
    \begin{tabular}{lll} % four columns, alignment for each
        \hline
         Parameter              &  Value$^{ \dagger}$  & Reference    \\
         \hline
          Age~(Myr)            & 625 (50)       & \citet{PBL98}. \\
          $d$~(pc)             & 47.51 (03)     & \citet{BRF21} \\
          $a~({R_{\star}})$    & 3.586 (11)     &  \citet{VWV15}  \\
          $M_\star~({M_{\sun}})$ &  0.9971 (12) &  \citet{VWV15}    \\
         $R_\star~({R_{\sun}}) $ & 0.93709 (93) &  \citet{VWV15} \\
         ${T_\mathrm{eff}}$~(K) & 5,066 (04)     &    \citet{VWV15} \\
         $\log g$~(cm/s) &  4.49331 (87)        &  \citet{VWV15}\\
         $P_\mathrm{rot} = P_\mathrm{orb}$~(d) & 0.5211833875 (27)   &  \citet{VWV15}  \\
         $i~(\degr)$  &  78.755 (30)           & \citet{VWV15}  \\
         $v\sin(i)$~(km/s) & 89.30 (11)         & \citet{VWV15}  \\
        \hline
        \multicolumn{3}{l}{\footnotesize$^\dagger$    Standard error of the last two digits is shown inside the parenthesis.} \\
    \end{tabular}
\end{table}

Photometric studies reveal an apparent magnitude ranging from $V=9.30$ to $9.42$ for the K2 dwarf star \citep[cf. Fig.~6 in][]{VWV15}. Given the distance modulus of $-3.384\pm0.002$ and the V-band bolometric correction at the effective temperature of the K2 dwarf star, $BC_\mathrm{V}=-0.29\pm0.02$ \citep{PM13}, we estimate minimum and maximum bolometric magnitudes of $M_\mathrm{bol, min} =5.626 \pm 0.063$ and $M_\mathrm{bol, max} =5.746 \pm 0.063$, respectively. Additionally, using the radius and effective temperature listed in Table~\ref{tab:paramsv471}, as well the reference bolometric magnitude for the Sun, $M_\mathrm{bol, \sun}=4.74$, we infer the bolometric magnitude for the unspotted star, $M_\mathrm{bol, u} = 5.451 \pm 0.004$. The fraction of spots at the surface of the K2 dwarf star is then given by $f_\mathrm{spot} = 1 - 10^{\frac{2}{5}(M_\mathrm{bol, u}-M_\mathrm{bol})} \approx~0.15$--$0.25$ which falls within the observed range of magnitudes. This spottedness of 15--25~per cent is typical for active stars. Previous brightness reconstructions of the K2 dwarf star with Doppler imaging retrieved spot coverage of $\approx 0.20$ in 1992/1993 \citep{RHJ95} and 0.09 in 2002 \citep{HAS06}.

\begin{figure}
	\centering
	\includegraphics[width=0.5\textwidth]{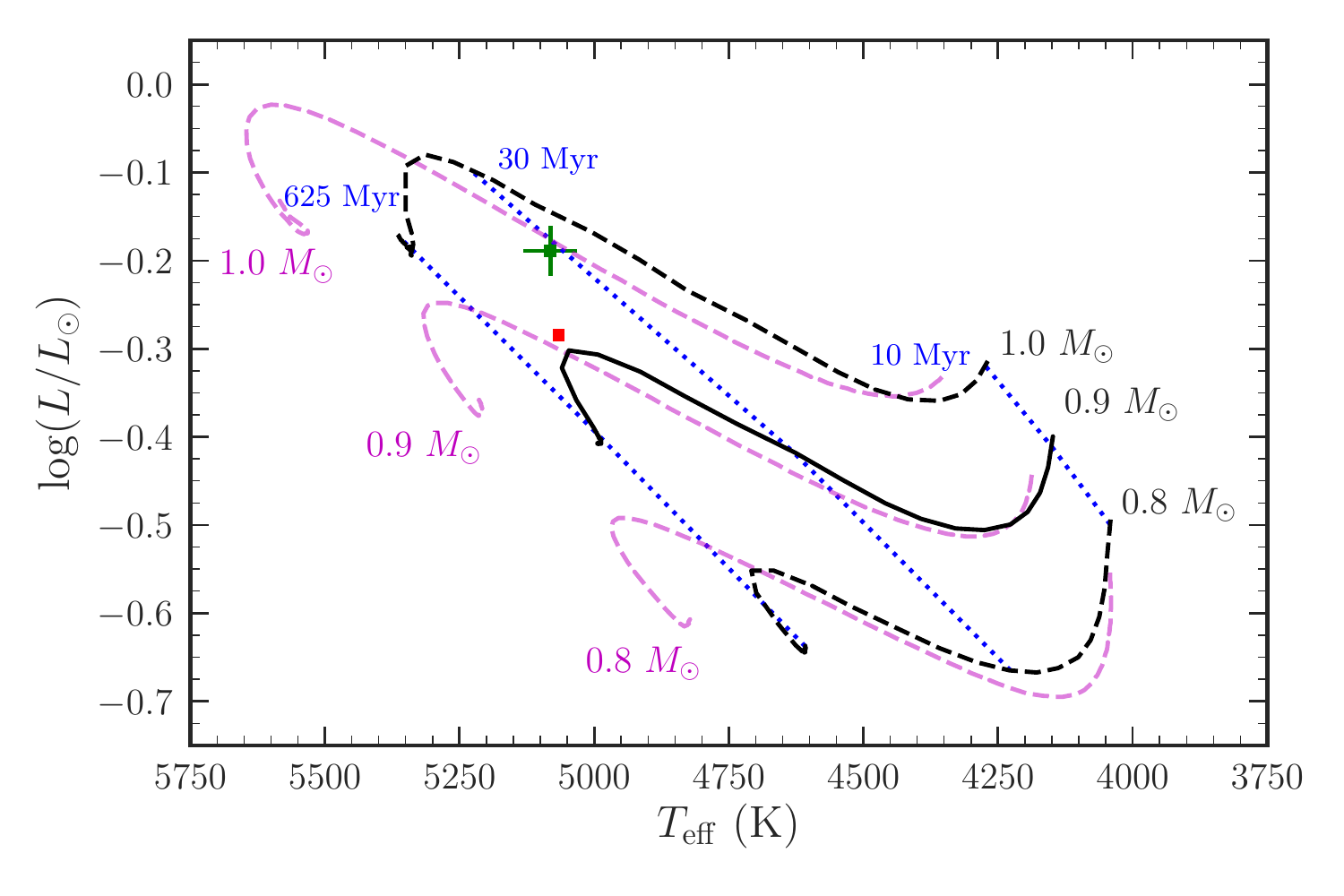}
    \caption{Positions of the K2 dwarf V471~Tau (red square) and the single-star analog AB~Dor\protect\footnotemark\
    (green square) in the Hertzsprung-Russell diagram. \citet{SDF00} evolutionary tracks for masses $0.8 - 1.0~M_\odot$ are shown in dashed black lines ($Z=0.0020$ + overshooting model), except the $0.9~M_\odot$ track shown in continuous black line, which we chose to represent the K2 dwarf star. Siess isochrones for $10$, $30$, and $625$~Myr are represented as dotted blue lines. Evolutionary tracks from \citet{BHA15} models (dashed magenta) are included for comparison.}
    \label{fig:evolK2}
\end{figure}

Figure~\ref{fig:evolK2} shows the K2 dwarf V471~Tau's position in the Hertzsprung-Russell diagram. Among the two evolutionary models considered, the 0.9~$M_\odot$ track of \citet{SDF00} is the one that best reproduces the stellar parameters of the K2 star. Still, we can notice the anomalous mass for its K2V spectral type that has been subject of investigations in the past years \citep[e.g.,][]{OBS01}. Some authors suggested that a metal enrichment during the common-envelope phase could potentially explain the over-mass; however, no conclusive answer exists yet \citep[see discussion in][]{VWV15}. Using the evolutionary model of  \citet{SDF00}, we infer that the radiative core of the K2 star reaches a radius of $0.68~R_\star$ at $625$~Myr (or, in other words, that it posses a convective envelope corresponding to the outer $32\%$ of the stellar radius).
\footnotetext{For the single-star analog AB~Dor, we adopt a luminosity based on the effective temperature of \citet{CTN07}, $5081 \pm 50$~K,  and the radius estimate of $1.05 \pm 0.10 R_{\sun}$ derived from Doppler imaging \citep{DCS03}, compatible with the estimate derived from interferometry \citep[$0.96 \pm 0.06 R_{\sun}$; ][]{GMM11}.}

%%%%%%%%%%%%%%%%%%%%%%%%%%%%%%%%%%%%%%%%%%%
\section{Observational data} \label{sec:obsdata}
%%%%%%%%%%%%%%%%%%%%%%%%%%%%%%%%%%%%%%%%%%%

We use spectropolarimetric data collected in two different seasons with ESPaDOnS at the Canada France Hawaii Telescope \citep{DCL06}. Our data set totalizes 230 unpolarised (Stokes $I$) and 56 circularly polarized (Stokes $V$) spectra spread in three non-consecutive days in 2004, and 400 unpolarised and 98 circularly polarized spectra spread in four days (separated by 1-d gaps) in 2005. Observations cover wavelengths from 370 to 1,000~nm at a resolving power of 65,000. We refer to \cite{D03} for further details about the Stokes parameters acquisition with ESPaDOnS. We note that circularly polarised spectra typically require 4 sub-exposures, obtained at different orientations of the polarimeter retarders, to be combined in an optimal way to minimize potential spurious signatures \citep[see][for more details]{DSC97}.

Raw data frames are reduced with the Libre-ESpRIT package, which is optimised for ESPaDOnS \citep{DSC97}.  The spectra collected in 2004 have peak signal-to-noise ratios (SNRs) ranging from 75 to 191 (median 147), while in 2005 it ranged from 75 to 188 (median 158). Circularly polarised spectra with peak SNRs lower than $75$ were rejected in this work, corresponding to four sequences in the first season of observation (Nov/Dec 2004) and three sequences in the second one (Dec 2005). The complete log of our observations can be found in Tables~\ref{table:data1} and \ref{table:data2} (Appendix~\ref{sec:tables}). We use the ephemeris of \citet{VWV15} to compute the rotational cycle $E$ of each observation\footnote{We follow the  convention proposed by \cite{B00} and express timings in Terrestrial Time (TT) scale. }
 \begin{equation} \label{eq:ephemeris}
 \mathrm{HJED} \qq{=}  2445821.898291 + 0.5211833875\times E,
\end{equation}
where phase 0.5 corresponds to the primary eclipse of the system (i.e., when the WD is in front of the K2 star).

\subsection{Least-squares deconvolved profiles}

Least-Squares Deconvolution  \citep[LSD,][]{DSC97} is used to produce an average profile of photospheric lines of the K2 dwarf star, with the SNRs boosted by a factor of 30 from the peak SNR of the individual spectra (see Tables~\ref{table:data1} and \ref{table:data2}) with respect to an average spectral line. We constructed the line mask using the Vienna Atomic Line Database \citep[VALD; ][]{PKR95,K93} for an effective temperature $T_\mathrm{eff} = $~5,000~K and a surface gravity $\log g = 4.5$, in agreement with \citet{VWV15} (see Table~\ref{tab:paramsv471}). We chose to include in our absorption line list only lines deeper than $10\%$ to the continuum level ($I_\mathrm{c}$), resulting in roughly 6,000 atomic lines. The average line profile features a mean wavelength $\lambda = 625$~nm, a mean relative depth $d = 0.677$, and a mean effective Land\'{e} factor $w = 1.2$.

%%%%%%%%%%%%%%%%%%%%%%%%%%%%%%%%%%%%%%%%%%%
\section{Zeeman-Doppler Imaging of V471~Tau} \label{sec:modelling}
%%%%%%%%%%%%%%%%%%%%%%%%%%%%%%%%%%%%%%%%%%%
We analyze the time series of the spectropolarimetric data using ZDI to obtain information on the brightness and magnetic field distributions at the surface of the K2 dwarf of V471 Tau. First introduced by \citet{S89}, ZDI traces how distortions in Stokes $I$ and $V$ profiles retrieve maps of the stellar surface.

We use ZDI as described in a suite of papers \citep{DSP89,BDR91,DB97}, using the implementation of \citet{D01} and adopting a spherical harmonic decomposition for the magnetic field. In short, ZDI decomposes the stellar surface in a grid of $N$ cells (N being typically 10,000). Synthetic profiles are computed locally in each cell, using the analytical solution of Unno-Rachkovsky to the polarised radiative transfer equations in a Milne-Eddington model of atmosphere \citep{LDL04}. Then, local profiles from each grid cell are Doppler-shifted according to the radial velocity (RV) of the cell position and weighted by a linear limb-darkening law. The local RV of each cell is related to the geometry of the system (see Sec.~\ref{Section:input}) and to the rotation profile assumed at the stellar surface (either a simple solid-body rotation or a Solar-like square-cosine-type latitudinal differential rotation, see Sec.~\ref{sec:diff}). These local profiles are then combined into global synthetic Stokes $I$ and $V$ line profiles that are directly compared to the time series of the Stokes $I$ and $V$ LSD profiles.

In the next step, maximum-entropy principles are applied to both brightness and magnetic reconstructions \citep{SB84}. The code reconstructs surface maps with a conjugate gradient algorithm that searches for the lowest amount of information capable of fitting the data at a given $\chi^2$ level (similarly, one can minimize the $\chi^2$ at given information content using an iterative procedure). For brightness reconstructions, the principle is directly applied to the local brightness of the grid cells, while in magnetic reconstructions, the entropy is a function of the spherical harmonics coefficients \citep{DHJ06}. In this study, we truncate the spherical harmonic representation of the magnetic field at order $\ell=15$, which is enough to extract most spatial information available in the line profiles.

\subsection{System parameters}\label{Section:input}
We take advantage of the maximum-entropy fitting process to simultaneously estimate the orbital parameters describing the RV of the K2 dwarf of the binary system;  since the orbit of V471~Tau is circular, there are 3 such parameters, the semi-amplitude of the orbital motion of the K2 dwarf ($\parx$), the systemic velocity ($\pary$), and the phase offset with respect to the ephemeris of Eq.~\ref{eq:ephemeris} ($\parz$). We perform a 3D search in the parameter space $\parx$, $\pary$, and $\parz$ to find out how $\chi^2$ varies (at constant reconstructed information at the surface of the star) with these parameters. By fitting a 3D paraboloid around the minimum of the derived $\chi^2$ values, we compute the best estimates of the parameters and their uncertainties. An inspection of these parameters shows that slightly different systemic velocities (by about $3\sigma$) minimize phase-coherent patterns present in the residuals (observed minus modelled Stokes profiles) at both epochs. We adopt then $\parx = 149.3 \pm 0.2$~km/s and $\pary = 35.0 \pm 0.1$~km/s. For the phase offset, we obtain for our Nov/Dec 2004 data set $\parz = 0.0040 \pm 0.0002$  and for the Dec 2005 data $\parz = 0.0035 \pm 0.0002$ (here a positive value in the phase offset, $\parz > 0$, indicates a later conjunction when compared to the prediction from ephemeris in Eq.~\ref{eq:ephemeris}). Both values of $\parz$ agree within the error bars.

Likewise, once the data are corrected for the orbital motion, we search for the projected rotational velocity $v\sin(i)$ that allows our synthetic profiles to match best the times series of Stokes $I$ LSD profiles. For both data sets, the line-of-sight projected equatorial rotation velocity associated with the lowest $\rchi$ is consistent within $2.5\sigma$ with \citet{VWV15}, i.e., $v\sin(i) = 89.30\pm 0.11$~km/s.

%%%%%%%%%%%%%%%%%%%%%%%%%%%%%%%%%%%%%%%%%
\section{Surface maps} \label{sec:maps}
%%%%%%%%%%%%%%%%%%%%%%%%%%%%%%%%%%%%%%%%%
We carry out reconstructions of brightness and magnetic maps of the K2 dwarf of V471~Tau for both data sets using the orbital and stellar parameters obtained in the previous section. As we discuss in Sec.~\ref{sec:diff}, differential rotation is detected at the surface of the K2 dwarf star, and we take it into account in the imaging process.

\subsection{Brightness maps}
\begin{figure*}
	\centering
	\includegraphics[width=\textwidth]{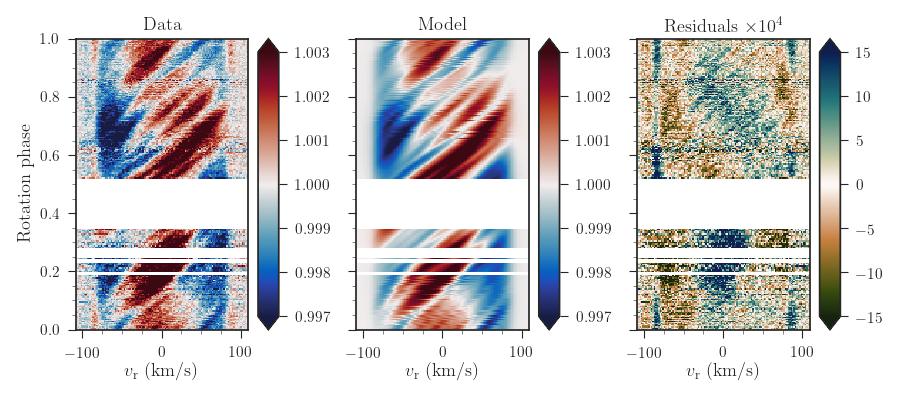}
	\includegraphics[width=\textwidth]{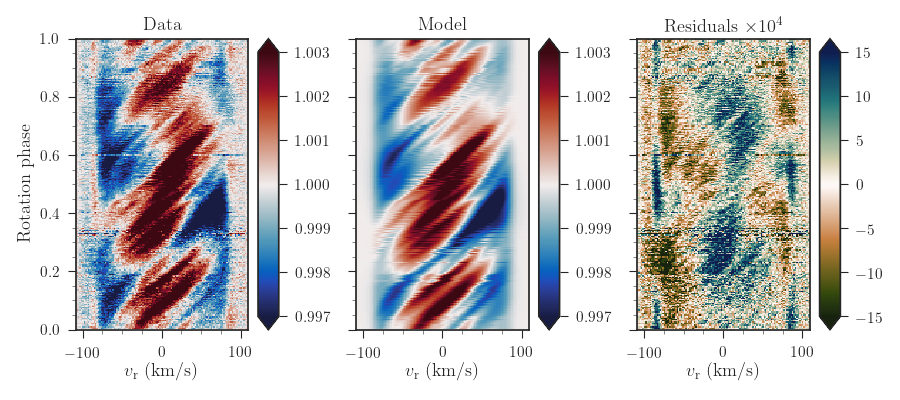}
    \caption{Dynamical spectra of the intensity flux (Stokes $I$) of the K2 dwarf of V471 Tau for our Nov/Dec 2004 (top) and Dec 2005 (bottom) data sets. From left to right, we show the observed LSD profiles, the modeled observations, and the residuals (i.e., observations minus model). LSD and modeled spectra were divided by the synthetic line profile of an unspotted star to emphasize spot signatures. We multiplied the residuals by a factor of $10^4$ for display purposes.}
    \label{fig:dynamic}
\end{figure*}

Figure~\ref{fig:dynamic} shows the dynamical spectra of Stokes $I$ profiles (for the individual profiles see Appendix~\ref{sec:StokesAppendix}). These dynamical spectra exhibit obvious signatures crossing the spectral line from the blue wing to the red wing, generated by surface brightness features being carried across the visible hemisphere as the star rotates. The few low-level features still present in the residuals, for instance, the blue and red vertical bands located at $\pm v\sin{i}$, reflect the fact that the width of the LSD profile slightly varies with phase as a result of the finite integration time (blurring the spectral lines at conjunction phases, i.e., when they move fastest, and thereby making them slightly wider). However, the typical amplitude of these residuals ($\sim\negmedspace10^{-3}$) is low enough not to cause any significant spurious features in the reconstructed images. 

\begin{figure}
    \centering
    \includegraphics[width=.4\textwidth]{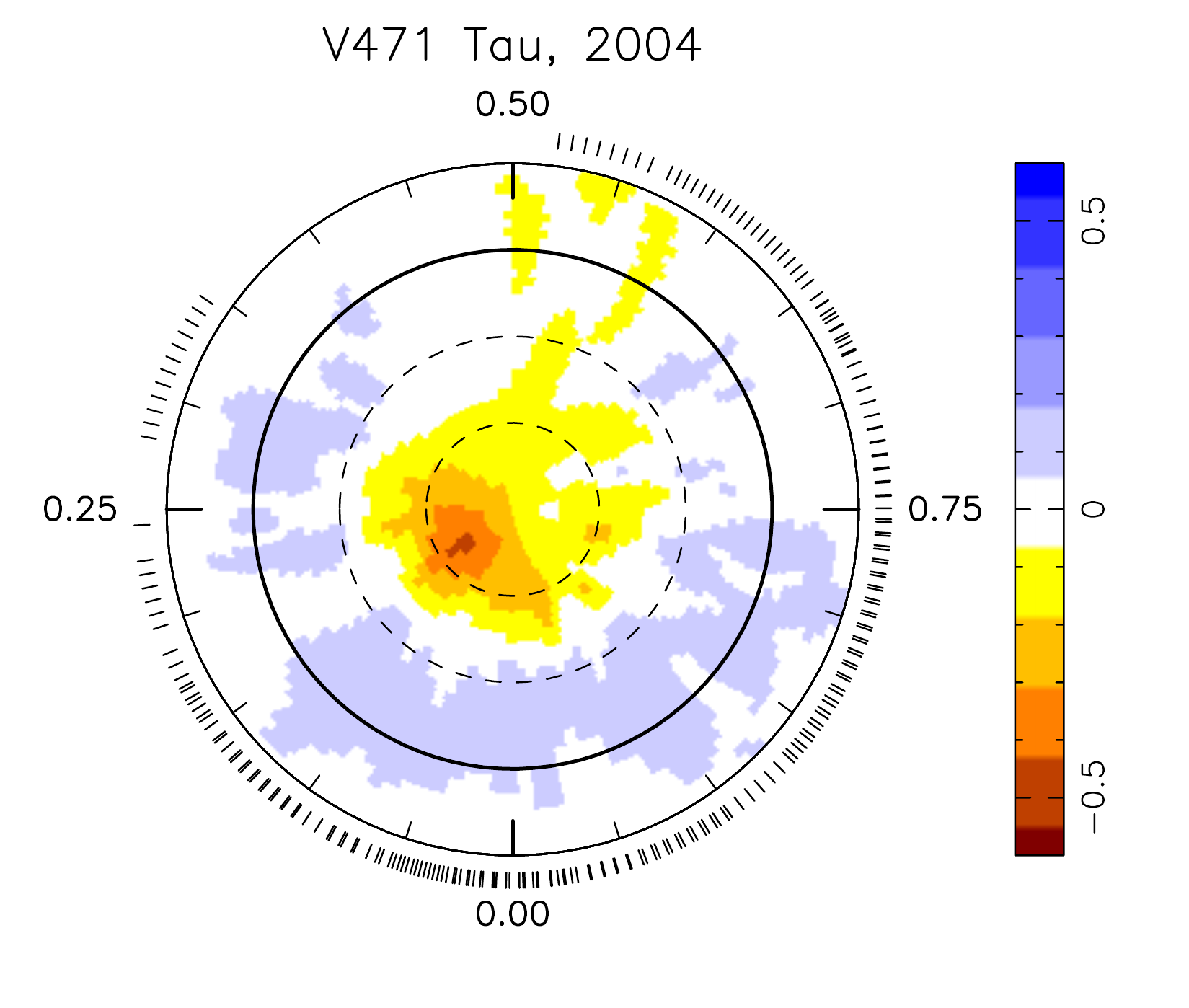} 
    \includegraphics[width=.4\textwidth]{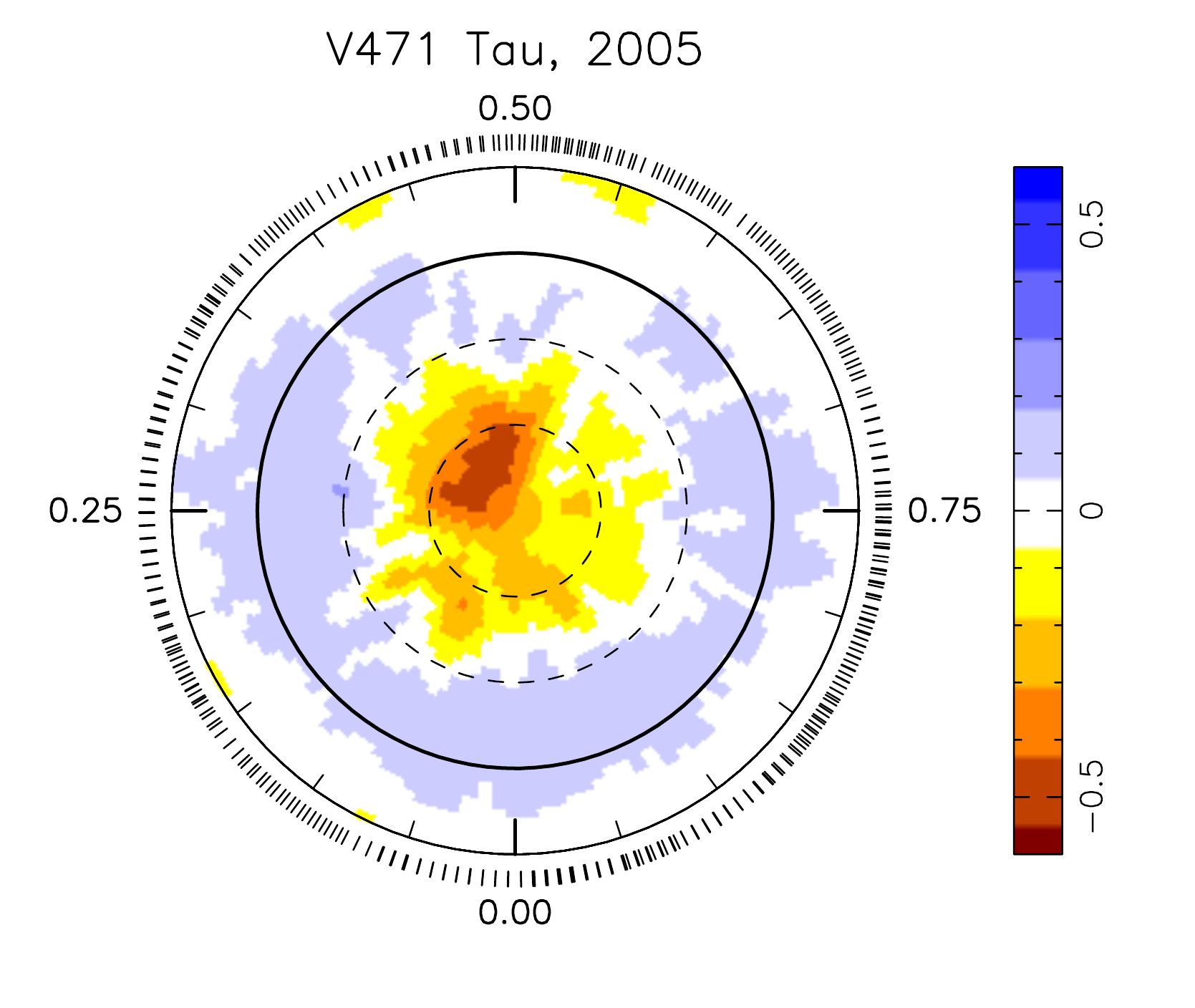}   
     \caption{Brightness maps obtained in Nov/Dec 2004 (top panel) and Dec 2005 (bottom panel). The logarithm of the relative brightness is shown in colors, with brown shades representing cool spots and blue shades depicting bright plages. In each polar representation, concentric circles denote regions of iso-latitude plotted in steps of $30\degr$ from the inner to the outermost circle. Ticks outside the polar representation indicate the rotation phase of the observations used to recover the maps.}
    \label{fig:brightness}
\end{figure}

The reconstructed brightness maps are shown in Figure~\ref{fig:brightness}. In 2004, the spot distribution exhibited a cool polar spot off-centered towards phase 0.15, extending down to a colatitude of $\sim\negmedspace50\degr$. We likewise identify in 2005 a cool polar cap, although it presents a higher contrast with the quiet photosphere and is now off-centered towards phase 0.35. Besides, both spot maps show a partial ring of low-contrast warm features encircling the polar region with a latitudinal extension of $\sim\negmedspace40\degr$. The ring disrupts at phase 0.5, corresponding to the face turned to the WD companion. Overall, the polar spot distributions exhibit similar structures but with a phase shift of $0.20 \pm 0.05$ from 2004 to 2005 (a result also noticeable in the dynamical spectra of both epochs Fig.~\ref{fig:dynamic}). We find that, in 2004, cool spots and warm plages covered $8\%$ and $6\%$ of the stellar surface, respectively, whereas, in 2005, the spot coverage was $10\%$ for cool spots and $7\%$ for warm plages. The increase in spot coverage between the two epochs possibly relates to the denser phase coverage of our 2005 data.  

\subsection{Magnetic topology} \label{sec:magtop}
\begin{figure*}
    \centering
    \includegraphics[width=\textwidth]{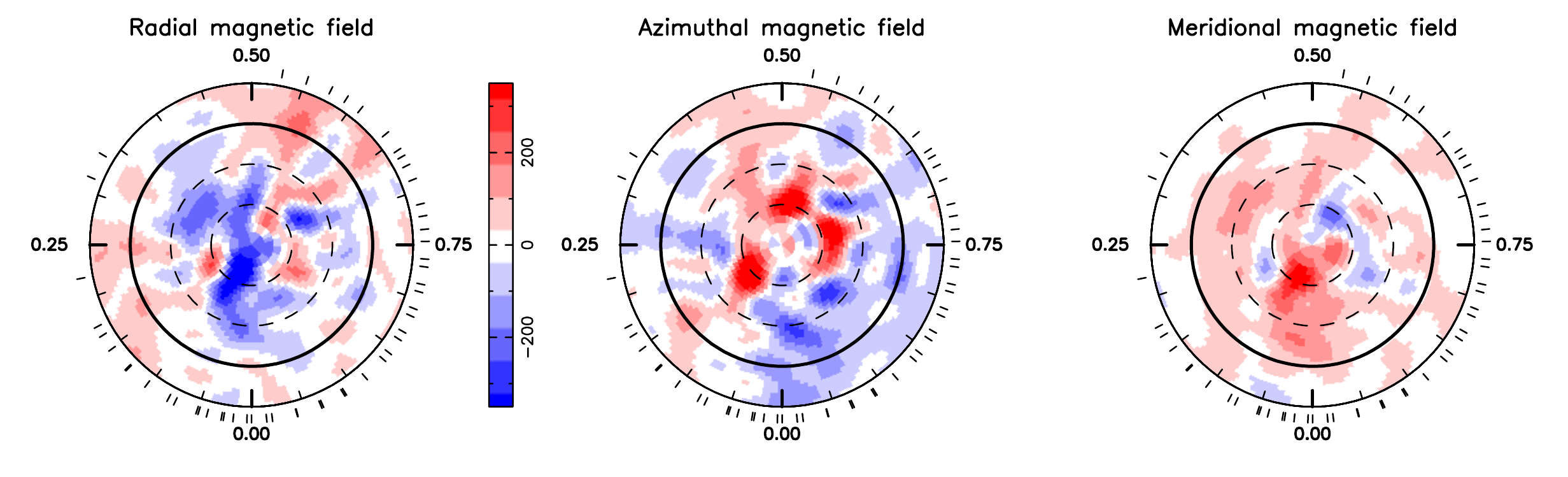} 
    \includegraphics[width=\textwidth]{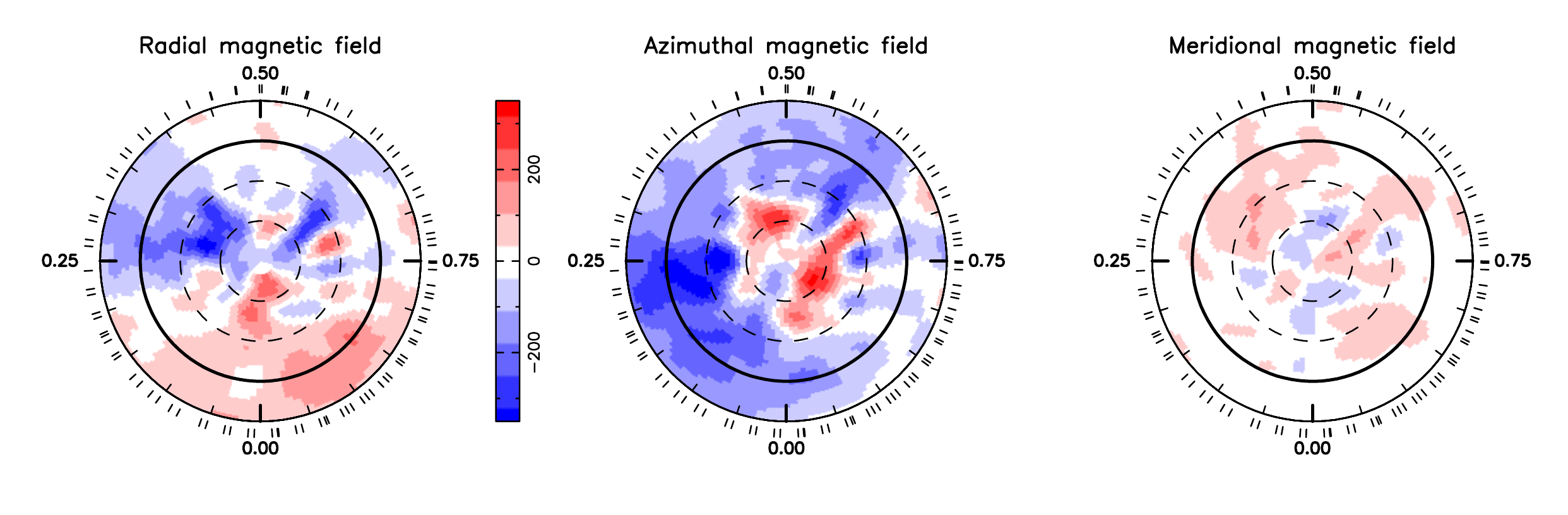} 
    \caption{Polar view of magnetic field topology for Nov/Dec 2004 (top panels) and Dec 2005 (bottom panels). From left to right, the columns show respectively the radial, azimuthal (i.e., toroidal), and meridional components of the large-scale magnetic field (with concentric circles and ticks as in Fig.~\ref{fig:brightness}). Magnetic field strengths are saturated at $350$~G, with red shades representing positive values and blue shades negative values.}
    \label{fig:magtop}
\end{figure*}

Observed Stokes $V$ profiles, with the ZDI fit down to a unit reduced $\chi^2$ level, are shown in Appendix~\ref{sec:StokesAppendix} (Figs.~\ref{fig:sv2004} and \ref{fig:sv2005}). The topology of the K dwarf's large-scale magnetic field is depicted in Fig.~\ref{fig:magtop}. We found a maximum radial field strength of 250~G in 2004 and 230~G in 2005, while the root-mean-square magnetic field was $\sim\negmedspace160$~G. In Nov/Dec 2004, the magnetic field topology shows a complex configuration with 60$\%$ of the magnetic energy reconstructed in modes with $\ell \geq 4$, whereas, in 2005, the energy stored in these modes dropped to 30$\%$. The remaining energy is mostly stored in the dipolar ($\ell = 1$) and octupolar ($\ell = 3$) components, but  quadrupolar components totalling up to $\sim\negmedspace10\%$ of the magnetic energy are also present in the reconstructed topology. In 2004, the magnetic field featured a $-90$~G dipole tilted by $20\degr $ to the rotation axis towards phase $0.08 \pm 0.03$. In 2005, the intensity of the dipole component was $-105$~G and the $64\degr$ tilt goes towards phase $0.41 \pm 0.03$. To assess the uncertainties of the image reconstruction, we performed 120 magnetic inversions at each epoch from bootstrapped data sets constructed by randomly choosing spectra from the original data, allowing for duplicates to match the original size of the sample \citep[e.g., see ][]{WSC17,WSG18}. Table~\ref{mag_tab} lists the main properties of the reconstructed large-scale magnetic topology along with the standard deviations obtained in the bootstrapping analysis. 

\begin{table}
\addtolength{\tabcolsep}{-3pt}
	\centering
	\caption{Magnetic field proprieties of the K dwarf star at Nov/Dec 2004 and Dec 2005. B$_\mathrm{rms}$ is the root-mean-square field, B$_\mathrm{dip}$ is the dipolar strength, and E$_\mathrm{pol}$ is the fractional energy in the poloidal field. E$_{\ell = 1}$, E$_{\ell = 2}$, E$_{\ell = 3}$ and E$_{\ell \geq 4}$ are, respectively, the fractional energies of the dipolar, quadrupolar, octupolar, and multipolar (defined as $\ell \geq 4$) components.}
	\label{mag_tab}
	\begin{tabular}{lcccccccc} 
	          \hline
 Date &  B$_\mathrm{rms}$ & B$_\mathrm{dip}$& $\theta_\mathrm{dip}$ & E$_\mathrm{pol}$ & E$_{\ell = 1}$  &  E$_{\ell = 2}$  & E$_{\ell = 3}$  &  E$_{\ell \geq 4 }$   \\
  & (G) & (G) & $\left(\degr \right)$ & $(\%)$ & $(\%)$ & $(\%)$ & $(\%)$ & $(\%)$ \\
 \hline
 2004 & $160 \pm 3$ & $-90 \pm 20$& $ 20 \pm 10$ & $ 70 \pm 5$ & $ 15 \pm 6$ & $ 10 \pm 2$ & $ 15 \pm 2$ & $ 60 \pm 7$ \\
 2005 & $160 \pm 1$ & $-105 \pm 5$ & $ 64 \pm 5$ & $ 60 \pm 2$ & $ 45 \pm 3$ & $ 10 \pm 1$ & $ 15 \pm 2$ & $ 30 \pm 2$\\
	         \hline		
	\end{tabular}
\end{table}

\section{Surface differential rotation}\label{sec:diff}

Thanks to the ability of ZDI to recover spatial information from sets of phase-resolved spectropolarimetric observations,  it is possible to retrieve information on differential rotation at the star's surface by finding out the recurrence rates of reconstructed features as a function of latitude. The procedure we use here, first described by \citet{DMC00}, takes into account an \textit{a priori} dependence of the angular velocity with latitude in the image reconstruction process. We adopt a Sun-like differential rotation law, written as:
\begin{equation}
    \Omega(\theta) = \Oeq -\dO \cos^2(\theta),
\end{equation}
where $\theta$ is the colatitude, $\Oeq$ is the angular velocity at the equator, and $\dO$ is the difference between $\Oeq$ and the angular velocity at the pole. Because this functional form depends on two free parameters, the reconstructed tomography likewise relies on the choice of $\Oeq$ and $\dO$. The differential rotation parameters, therefore, correspond to the set of parameters leading to the images with the lowest information content (similar to the procedure outlined in Sec.~\ref{Section:input}).  

\begin{figure*}
        \centering
	\includegraphics[width=.47\textwidth]{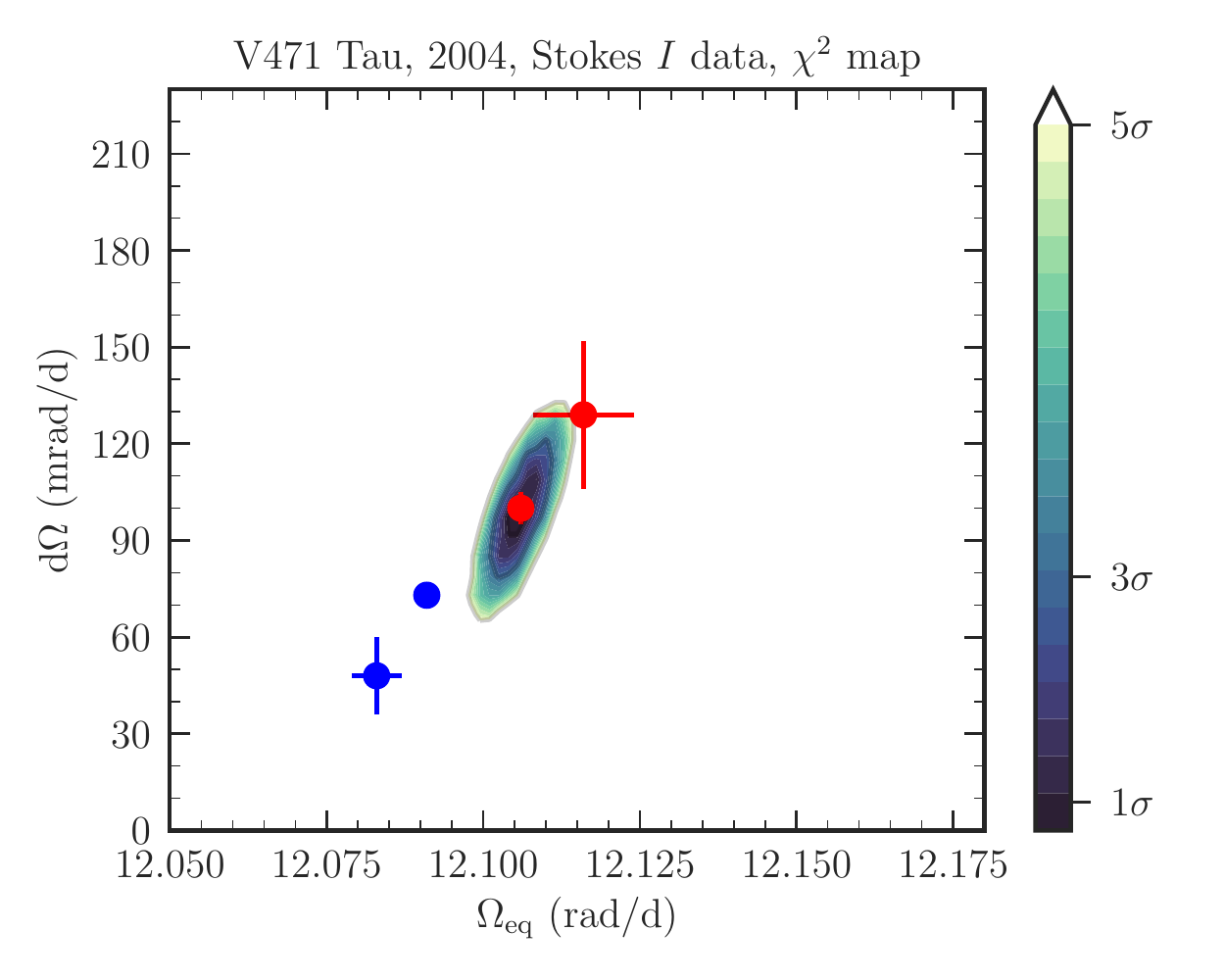}	
	\includegraphics[width=.47\textwidth]{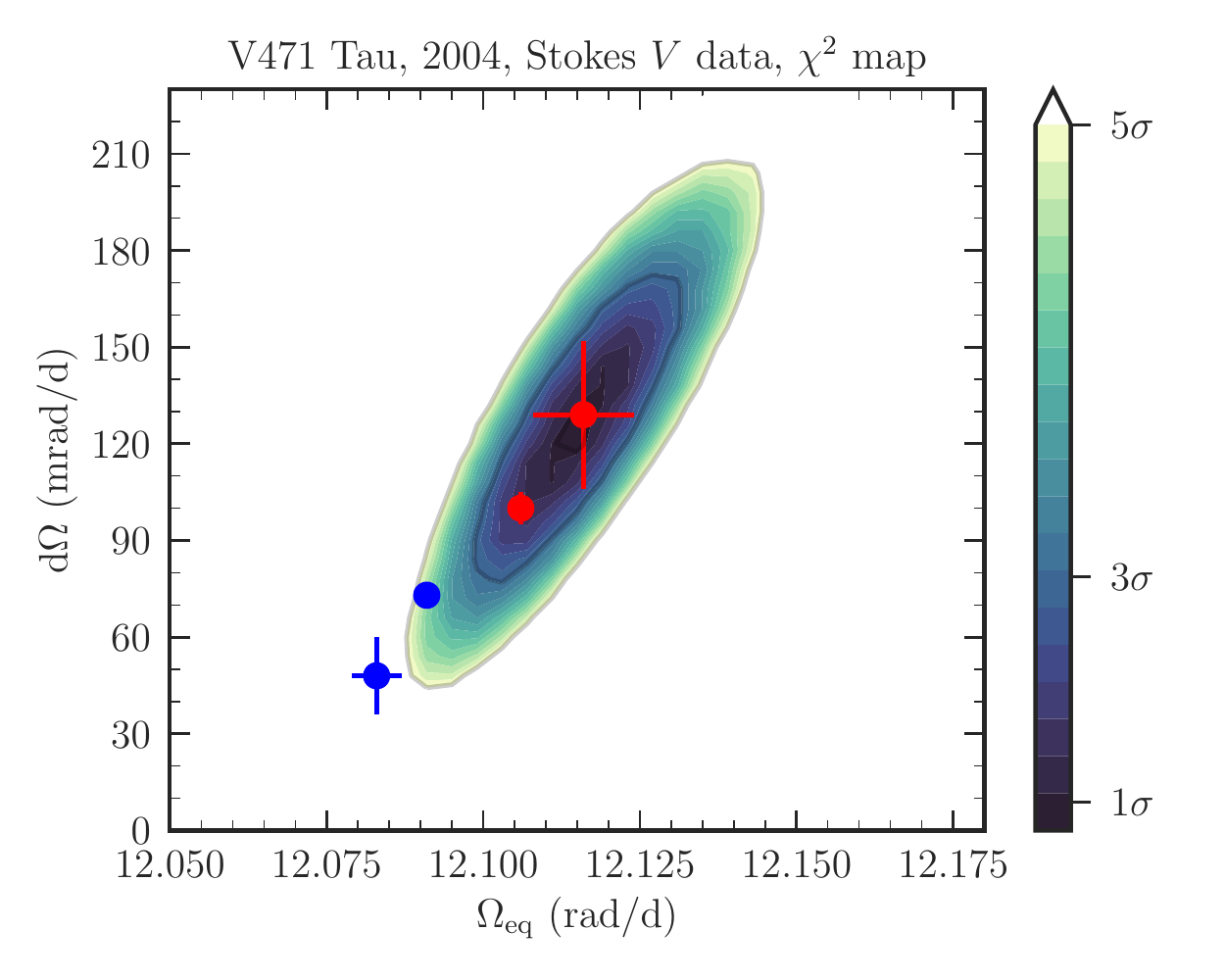} \\ 
	\includegraphics[width=.47\textwidth]{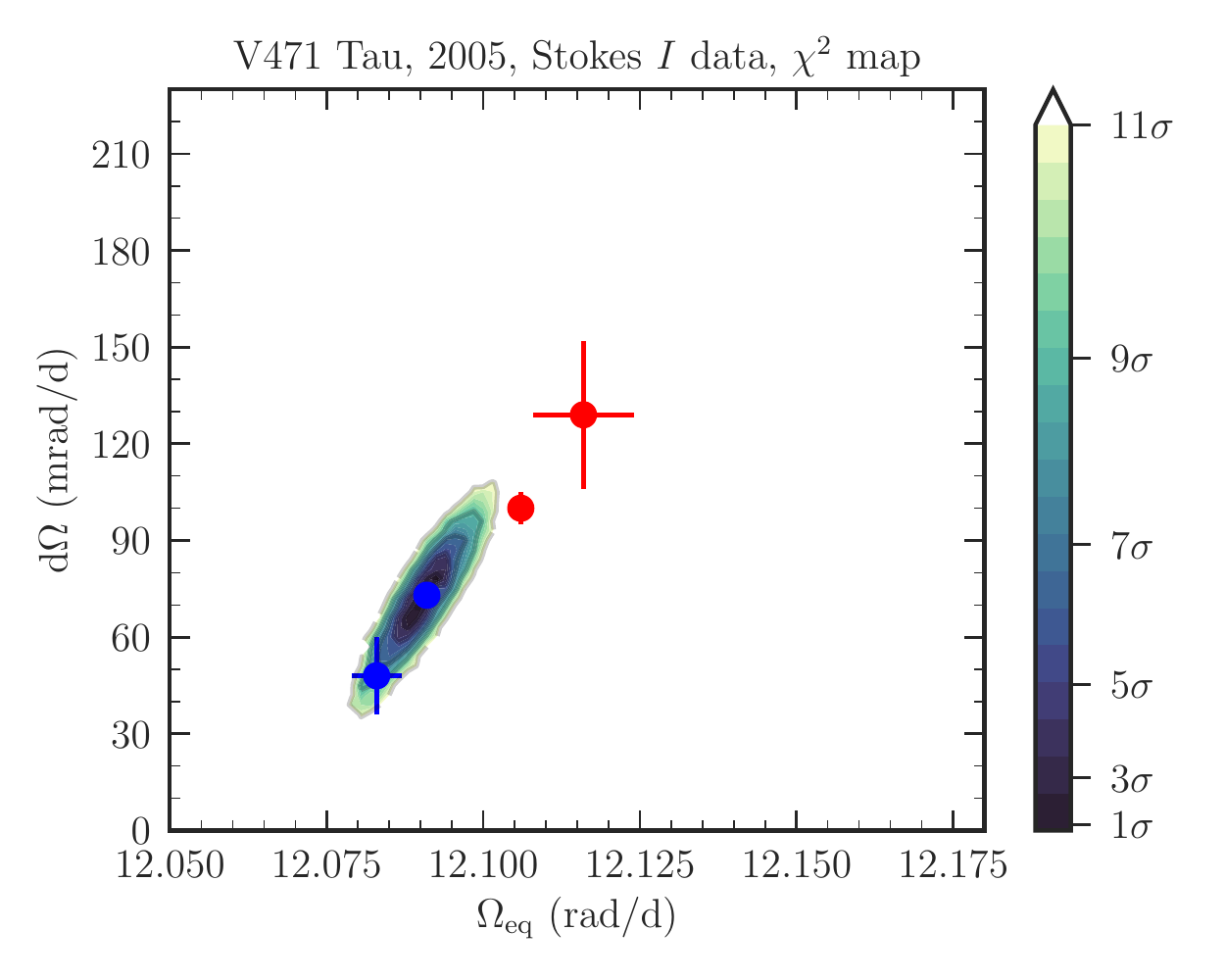}
	\includegraphics[width=.47\textwidth]{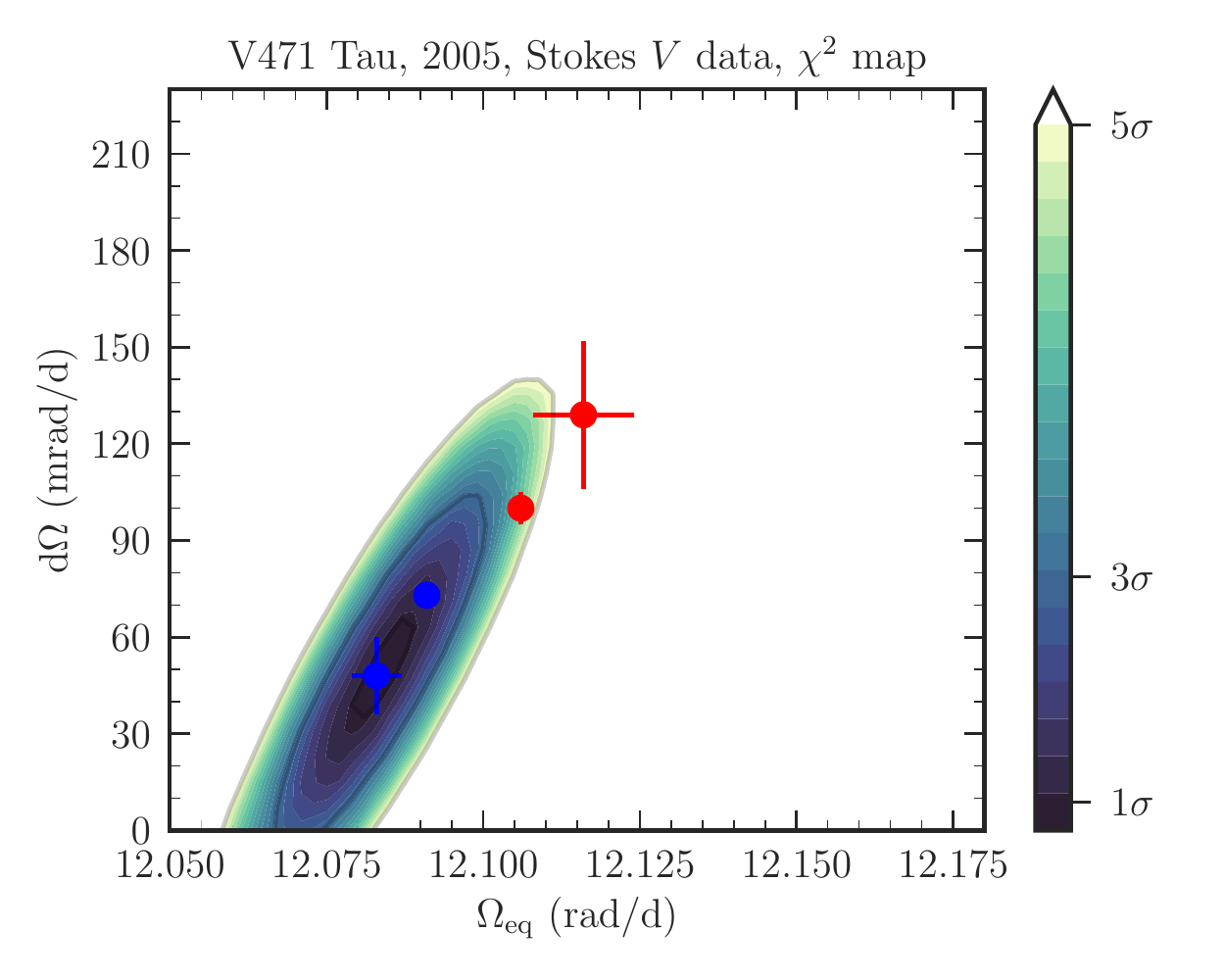}
    \caption{$\rchi$ maps in the differential rotation plane obtained with reconstructions of the brightness distribution (left panels) and with reconstructions of the magnetic topology (right panels). Top plots show the results for Nov/Dec 2004 data and bottom ones for Dec 2005 data. Confidence levels up to 5~$\sigma$ (11~$\sigma$ for Stokes $I$ data collected in 2005) are shown in colors and are computed with respect to the $\rchi$ minima. The circles and 1-$\sigma$ error bars indicate the center of the $\rchi$ distribution given by Eqs. 2 and 3 in \citet[][]{DCP03}, with the measurements of 2004 shown in red and those of 2005 in blue. Note that we repeated the 4 differential rotation measurements in each panel to aid comparisons.}
    \label{fig:dr}
\end{figure*}

Figure~\ref{fig:dr} shows $\chi^2$ maps in the $\Oeq$ -- $\dO$ plane for reconstructions using Stokes $I$ (left column) and Stokes $V$ (right column) signatures. We again assume a simple paraboloid approximation for the $\rchi$ maps close to the minimum to determine the optimal differential rotation parameters (circles) and their associated error bars \citep[see Eqs. 2 and 3 in ][]{DCP03}. We summarise the differential parameters found for each case in Table~\ref{diff_table}. The four detection reveal equatorial regions spinning faster than the polar ones, which corresponds to a solar-like shear. We additionally list in Table~\ref{diff_table} the colatitude $\theta_\mathrm{c}$ at which the system rotates with the orbital period and the colatitude and rotation rate of the spots' gravity center ($\theta_\mathrm{s}$ and $\Omega_\mathrm{s}$).

For an independent check that the error bars on the differential rotation parameters we derived are reliable, we constructed ten bootstrapped data sets for both epochs of observation (Nov/Dec 2004 and Dec 2005). We repeated the procedure described in this section for the bootstrapped data sets and derived ten differential rotation measurements for each Stokes profile. We find that the mean values of $\Oeq$ and $\dO$ agree within error bars with the values quoted in Table~\ref{diff_table} and that the dispersion on these two parameters is consistent with the error bars derived from the $\chi^2$ maps. Moreover, the mean value of the error bars derived from the $\chi^2$ maps is similar to those quoted in Table~\ref{diff_table}.

\begin{table*}
	\centering
	\caption{Differential rotation parameters derived from our Nov/Dec 2004 and Dec 2005 observations. Columns 2 - 7 show results obtained from Stokes $I$ data and columns 8 - 13 those obtained with Stokes $V$ data. Equatorial rates $\Oeq$ are listed in columns 2 and 8, while differential rotation rates $\dO$ are shown in columns 3 and 9 along with $1\sigma$ error bars for both quantities. For future reference, we also provide the colatitude of co-rotation ($\theta_\mathrm{c}$), the colatitude of the gravity centre of the spot/magnetic distribution \citep[$\theta_\mathrm{s}$;][]{DMC00}, and the rotation rate at this colatitude ($\Omega_\mathrm{s}$). The number of data points ($n$) used in each image reconstruction is provided in columns 7 and 13. }
	\label{diff_table}
	\begin{tabular}{lccccccccccccc} 
	          \hline
	                    & \multicolumn{6}{c}{Stokes $I$ / Brightness reconstruction}  & & \multicolumn{6}{c}{Stokes $V$ / Magnetic field reconstruction}       \\		
%	          \cline{2-3} \cline{5-6}
	       	 Epoch & $\Oeq$  & $\dO$  & $\theta_\mathrm{c}$ & $\theta_\mathrm{s}$ & $\Omega_\mathrm{s}$  & $n$ &  & $\Oeq$  & $\dO$   & $\theta_\mathrm{c}$ &  $\theta_\mathrm{s}$ & $\Omega_\mathrm{s}$ & $n$ \\
		            &   (rad/d)  & (mrad/d) &  ($\degr$)  & ($\degr$) &  (rad/d) &  &  &    (rad/d)    & (mrad/d)   &  ($\degr$)   &  ($\degr$)    &   (rad/d)        &     \\
		\hline
		 2004  & $ 12.106 \pm 0.001$   & $ 100 \pm 5 $ & $44 \pm 2$ &   $65$  & $12.088$ & 27572 &  & $ 12.116 \pm 0.008$   & $ 129 \pm 23 $ & $46 \pm 8$  &  $57$  & $12.078$ & 6344 \\
		 2005  & $ 12.091 \pm 0.001$   & $ 73 \pm 2 $ & $46 \pm 1$ &   $57$  & $12.069$ & 48800 &  & $ 12.083 \pm 0.004$   & $ 48 \pm 12 $ & $40 \pm 15$ &  $59$  & $12.070$ & 11590 \\
	         \hline		
	\end{tabular}
\end{table*}

%%%%%%%%%%%%%%%%%%%%%%%%%%%%%%%%%%%%%%%%
\section{Balmer lines} \label{sec:halpha}
\begin{figure}
    \centering
    \includegraphics[width=0.4\textwidth]{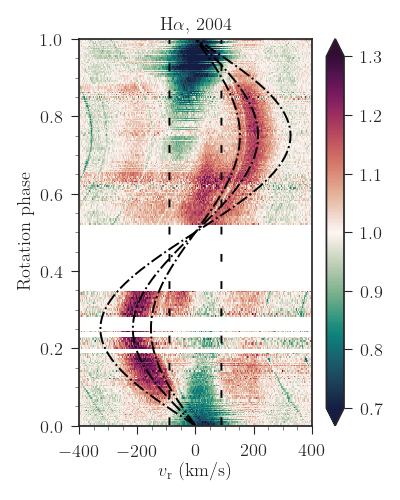} 
    \includegraphics[width=0.4\textwidth]{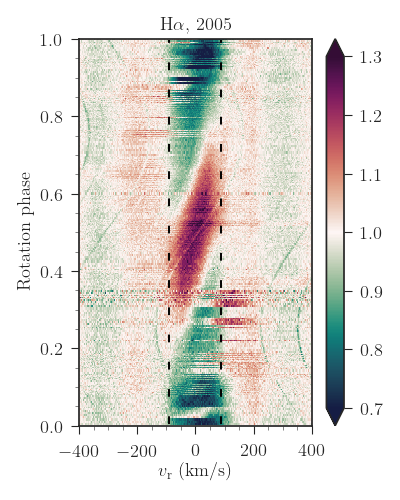}
    \caption{Dynamical spectra of H$\alpha$ line, in the rest frame of the K2 dwarf, for observations in late 2004 (top) and Dec 2005 (bottom). The vertical dashed lines correspond to the stellar rotational broadening of $\pm v\sin(i)$. Sine waves of amplitudes $150$~km/s (center of mass), $210$~km/s (prominence position), and $320$~km/s (WD position) are over-plotted on the 2004 dynamical spectrum. Weak features from telluric lines, noticeable by its sinusoidal behavior, remained after the removal procedure. }
    \label{fig:ha}
\end{figure}

%%%%%%%%%%%%%%%%%%%%%%%%%%%%%%%%%%%%%%%%
The H$\alpha$ line is often used as tracer of magnetic activity in stars. Figure~\ref{fig:ha} displays the dynamical spectra of H$\alpha$ for both observing epochs. We find that in both data sets, H$\alpha$ is modulated with orbital phase, being weakest at phase 1.0, i.e., when the WD is behind the K2 star, and strongest at phase 0.5, i.e., when the WD is in front of the K2 star. The equivalent width reveals a peak-to-peak modulation amplitude of about 1.6~{\AA} with a maximum emission reaching $-0.5$~{\AA} in Nov/Dec 2004 and $-0.6$~{\AA} in Dec 2005 (see Fig.~\ref{fig:haeq}). A similar modulation pattern is also visible in the core of other active lines (see Appendix~\ref{sec:lines}). 

By comparing the dynamical spectra of H$\alpha$ at both epochs, we detect an additional emission component moving in phase with the WD star in the 2004 data set. A Gaussian fit to the spectral signature around phase 0.75 yields an equivalent width of about $-0.33$~{\AA} and a full width at half maximum of  $1.95$~{\AA}. We interpret this extra emission as a prominence trapped in the magnetic field of the K2 dwarf. From the semi-amplitude of its signature in the dynamic spectrum ($210~\pm~38$~km/s in the rest frame of the K2 star), we can infer that the corresponding plasma is located at a distance of $2.35 \pm 0.43~\mathrm{R_\star}$ from the center of the K2 star towards the WD, i.e., at a distance of only $1.23~\mathrm{R_\star}$ from the WD. Although less obvious, a similar signature can also be identified in H$\beta$ after the removal of the stellar contribution (here assumed to be well represented by the prominence free spectra obtained in 2005 -- see Fig.~\ref{fig:activitylines}). Figure~\ref{fig:hb} shows the resulting spectra, where regions within $\pm v\sin{i}$ were masked to emphasize the prominence signature. From the isolated profile around phase 0.75, we estimate an H$\beta$ equivalent width of about $-0.096$~{\AA}, implying a Balmer H$\alpha$ to H$\beta$ decrement factor of 3.4.

\begin{figure}
    \centering
    \includegraphics[width=0.45\textwidth]{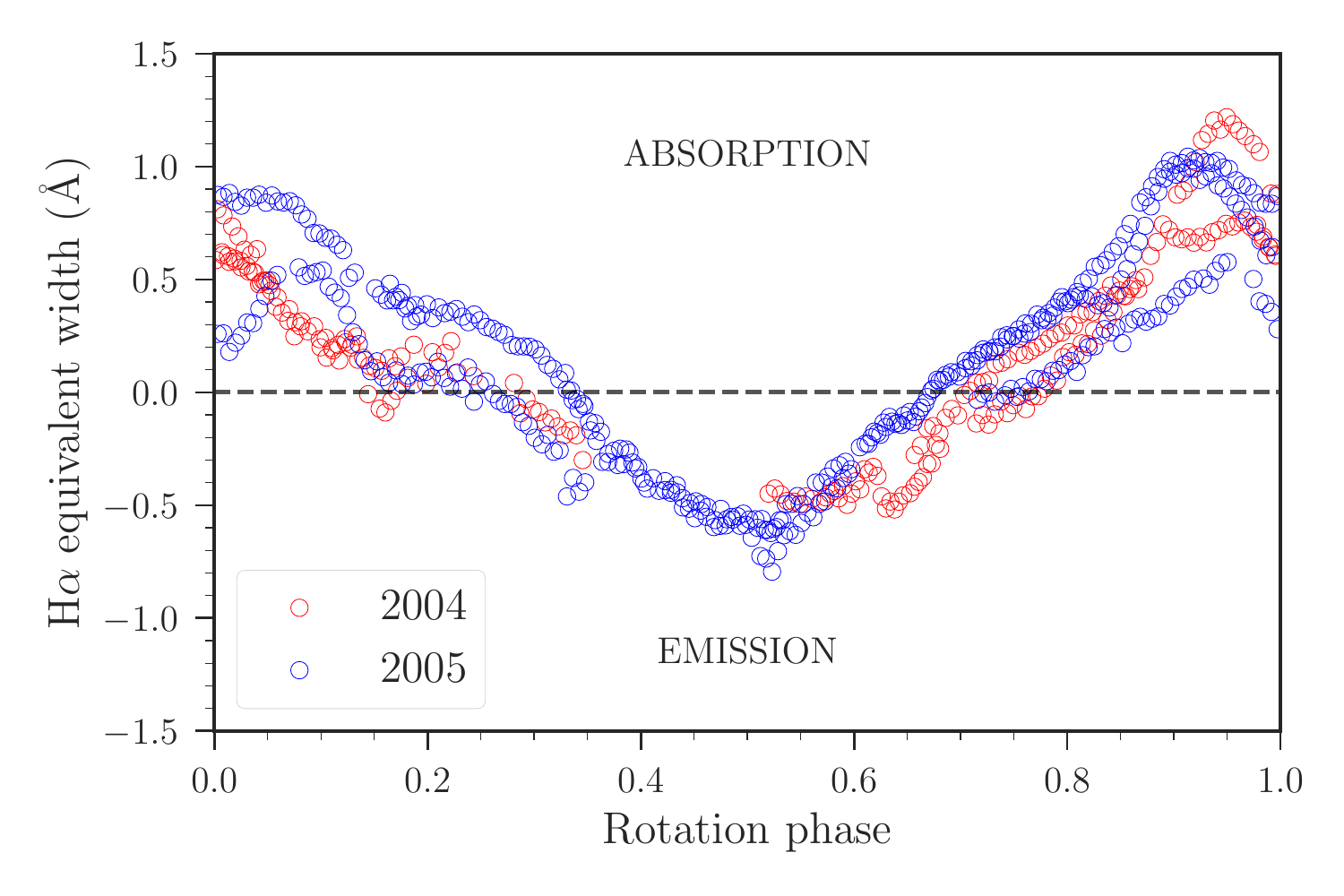}
        \caption{Rotational modulation in the H$\alpha$ equivalent width. Red symbols shown the data collected in Nov/Dec 2004 and the blue ones the data from Dec 2005.}
    \label{fig:haeq}
\end{figure}

\begin{figure}
    \centering
    \includegraphics[width=0.4\textwidth]{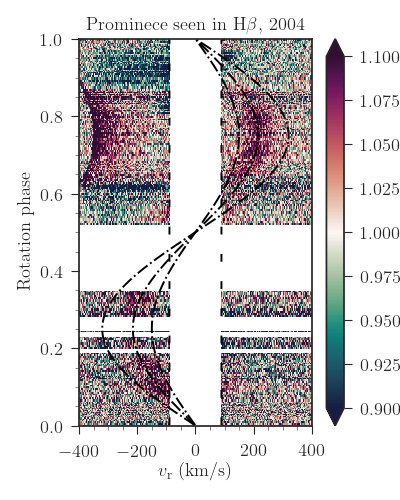}
        \caption{Dynamical spectra of H$\beta$ in 2004 after the removal of the stellar contribution (see text). Sine waves match the ones described in Fig.~\ref{fig:ha} and regions within $\pm v\sin{i}$ were masked for clarity.}
    \label{fig:hb}
\end{figure}

%%%%%%%%%%%%%%%%%%%%%%%%%%%%%%%%%%%%%%%%%%
\section{Discussion} \label{sec:disc}
%%%%%%%%%%%%%%%%%%%%%%%%%%%%%%%%%%%%%%%%%%
We analyzed spectropolarimetric data of the close-binary system V471~Tau acquired in Nov/Dec 2004 and Dec 2005. Photospheric lines of the K2 dwarf companion were identified after correcting for the orbital motion of the system (Sec.~\ref{Section:input}). We used ZDI to characterize the surface distribution of brightness and magnetic features at the surface of the K2 dwarf from the shapes and rotational modulation of Stokes $I$ and $V$ profiles. We discuss below the main results of this paper and how they contribute to the understanding of the various physical effects we aimed at studying.  

\subsection{Spot and magnetic structures}\label{sec:disc_maps}
We applied ZDI to our Nov/Dec 2004 and Dec 2005 data sets to reconstruct the brightness and magnetic maps at the surface of the K2 dwarf component of V471~Tau.  Our brightness maps exhibit a cool off-centered polar spot and a hot ring-like structure at low-latitude. The brightness maps we reconstruct resemble previous results obtained with similar techniques in 1992/1993 \citep{RHJ95} and 2002 \citep{HAS06}, especially at high latitudes. This implies that polar spots are long-lived at the surface of the K2 dwarf, as they are on the single-star analog AB~Dor \citep[e.g.,][]{DCS03}. The spot coverage we derive (14$\%$ and 17$\%$ in Nov/Dec 2004 and Dec 2005 respectively) are in good agreement with what is expected from photometry (in the range 15–25$\%$, see Sec.~\ref{sec:paramsv471tau}) suggesting that most of the brightness spots generating photometric fluctuations in V471~Tau are large enough to be detected and resolved by Doppler imaging.

The magnetic maps we obtained in this work are the first reconstructions achieved for the K2 dwarf V471~Tau (Fig.~\ref{fig:magtop}). The unsigned average magnetic flux at the surface of the star is $\sim\negmedspace200$~G, including a $\sim\negmedspace100$~G dipole component inclined at 20–60$\degr$ to the rotation axis. We note changes in the field topology between the 2 epochs, e.g., an increase in the strength of the toroidal component (from 30 to 40$\%$ of the reconstructed magnetic energy) and in the fractional energy of the dipolar component (from 15 to 45$\%$). However, we caution that such changes may at least partly reflect the improved phase coverage in our second observing session.

\subsection{Differential rotation and angular momentum distribution}\label{sec:fluctdiff} 
We detect differential rotation at the surface of the K2 dwarf at both epochs of observation (Table~\ref{diff_table}). In 2004, we found that the brightness and magnetic maps are sheared by $\dO=100\pm5$ and $129\pm23$~mrad/d, respectively; whereas, in 2005, these shears dropped to $\dO=73\pm2$ and $48\pm12$~mrad/d. These results differ from that of \citet{HAS06}, who found an almost solid body rotation ($\dO = 1.6\pm6$~mrad/d) for the star in 2002, already offering some tentative evidence for fluctuations in the surface shear on a short timescale ($\sim\negmedspace2$-yr). Furthermore, this finding reflects those of \citet{DCP03} who identified similar fluctuations in the single-star analog AB~Dor.

Our results show at both epochs that the magnetic topology suffers a different shear than the brightness distribution, which may reflect that brightness and magnetic features are anchored at different depths within the convective zone.  Following \citet{DCP03}, we propose to interpret the temporal fluctuations in the surface differential rotation of the K2 dwarf in terms of redistribution of angular momentum within the convective zone as the star progresses on its activity cycle.  Assuming angular momentum conservation in the convective zone, they found that variations in $\Oeq$ and $\dO$ should be correlated. For instance, in stars with a Sun-like angular rotation profile (varying with latitude and independent of radius), the correlation shows up as:
\begin{equation}
    \Oeq =  0.2 \dO + \Omega_0,
\end{equation}
where $\Omega_0=2\pi/P_\mathrm{rot}$. On rapid rotators however, where angular rotation is constant along cylinders according to Taylor-Proudman theorem, the correlation takes the following form: 
\begin{equation}
    \Oeq = \lambda \dO + \Omega_0,
\end{equation}
where $\lambda$ is a parameter related to the second and fourth-order moment of the fractional radius. For AB~Dor and the K2 component of V471 Tau, $\lambda$ is expected to be about $0.52$ \citep{DCP03}.

\begin{figure}
	\centering
	\includegraphics[width=0.46\textwidth]{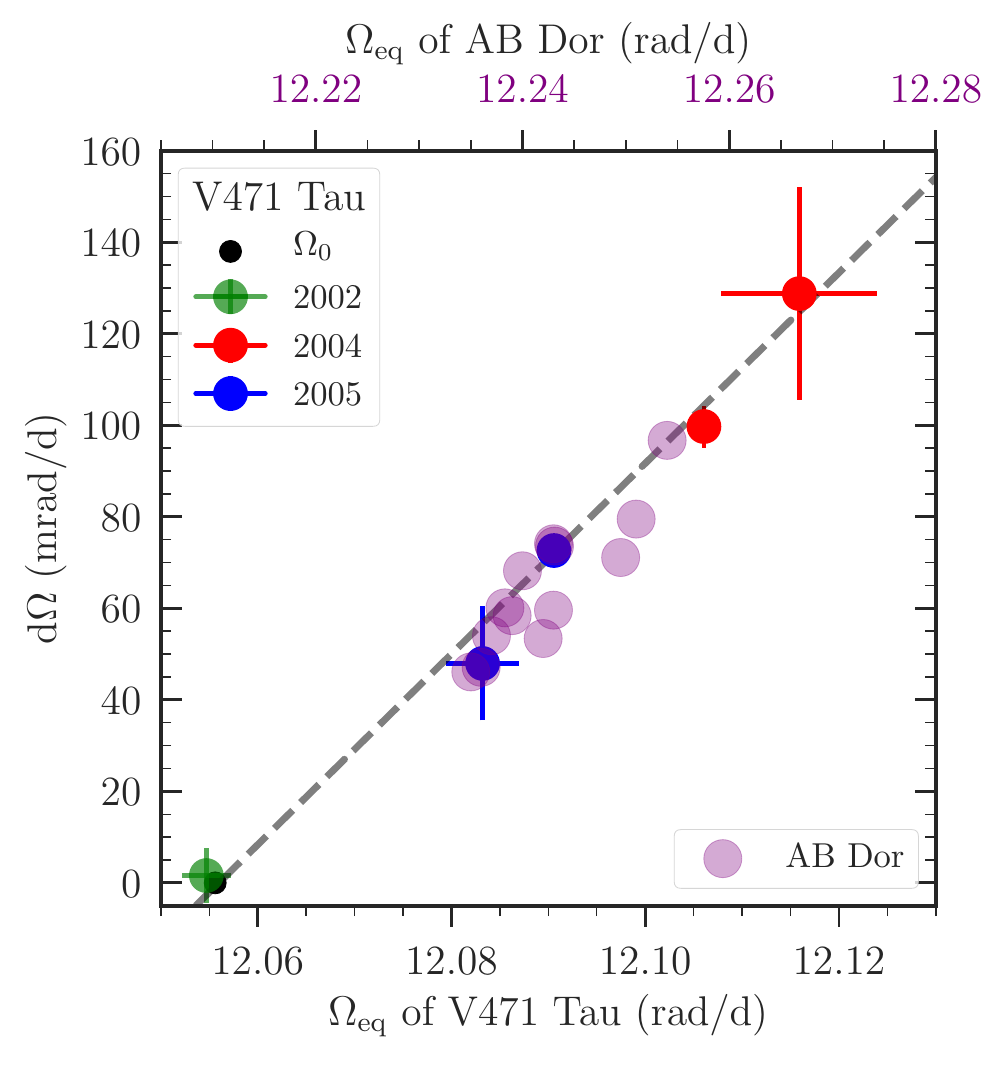} 
    \caption{Differential rotation at the K2 dwarf's surface obtained with our Nov/Dec 2004 (red) and Dec 2005 (blue) data sets with $\pm 1$–$\sigma$ level provided. The measurement of \citet{HAS06} in 2002 ($\Oeq=12.0547\pm0.001$~rad/d and $\dO = 1.6\pm6$~mrad/d) is the green cross and the angular velocity if the star was rotating as a solid body at the orbital angular velocity of the binary system, i.e., $\Omega = 2\pi/P_\mathrm{orb}$, the black dot.  A linear fit of these six quantities (dashed line) returns the following relationship  $\Oeq = (0.48 \pm 0.02) \dO + (12.056 \pm 0.002)$~rad/d. For comparison, we include shear measurements reported for the analog AB~Dor \citep[purple circles;][]{DCP03,JDC07}, after scaling the x-axis to the same rotation rate so that measurements for both stars can be compared.}
    \label{fig:angmom}
\end{figure} 

Figure~\ref{fig:angmom} shows the various existing estimates of differential rotation obtained so far for the K2 dwarf of V471~Tau. The linear fit to these values (in rad/d), including a solid-body rotation at the orbital period, yields the following trend: 
\begin{equation} \label{eq:fitdiff}
\Oeq = (0.48 \pm 0.02) \dO + (12.056 \pm 0.002)~\mathrm{rad/d}.
\end{equation} 
The slope we get, $\lambda = 0.48 \pm 0.02$, is consistent with expectations that rotation is constant on cylinders in stars rotating as fast as the K2 dwarf, as expected from the Taylor-Proudman theorem. 

In previous theoretical studies on close binaries, tides were claimed to be capable of quenching surface differential rotation \cite[e.g.,][]{S81,S82}. However, the shears reported in this work, as well as those of the HD~155555 binary system \citep[][]{DHC08}, do not confirm this conclusion. In particular, our result indicates that surface differential rotation in close binary stars is not specific to young stars like HD~155555. We also note that temporal fluctuations in the surface differential rotation of the K2 dwarf of V471~Tau tend to be larger than those reported for AB~Dor \citep[see purple circles in Fig.~\ref{fig:angmom};][]{DCP03,JDC07}, which may reflect the impact of tidal forces on dynamo processes. Nevertheless, it is noteworthy that in both stars shear variations follow the same trend in the $\Oeq - \dO$ plane (Fig.~\ref{fig:angmom}), which is further evidence in favor of our interpretation that angular rotation is constant along cylinders in the convective zone of these two similar active stars.

\subsection{Origin of ETV in V471~Tau}
We consider the different models proposed to explain the ETVs observed in the V471~Tau system in light of our results. Whereas the existence of a third body to explain ETVs is not completely ruled out, its existence currently seems unlikely \citep{HSP15}.

In the \citet{A92} mechanism, period modulations are an outcome of a cyclic redistribution of angular momentum induced by a dynamo mechanism operating within the active companion. The exchange of angular momentum throughout the activity cycle affects the star's oblateness, causing a modulation in the quadrupole moment and, therefore, changing the gravity in the orbital plane. When the quadrupole moment of the K2 star increases, the WD approaches the companion and the system's orbital period decreases to conserve angular momentum (and vice versa). 

Several authors questioned whether the Applegate mechanism could explain the ETVs of close binaries  \citep{L05,L06,VSB18}, since the cyclic exchange of angular momentum required in this model demands large shear fluctuations to explain typical period modulations \citep[e.g., $\Delta P/P_\mathrm{orb}\simeq 8.5 \times 10^{-7}$ for V471~Tau, cf.][]{L20}. In particular, for post-common envelope binary systems in which the active companion has a radiative core, \citet{VSB18} found that a relative differential rotation $\dO/\Omega \lesssim 1\%$ (compatible with our results range $0.4-1.1$ per cent) can only lead to period variations $\lesssim\negmedspace10^{-7}$, thus making it unlikely to occur in V471~Tau. However, because our observations were undertaken during a phase in the ETV cycle when the observed minus predicted eclipse timings are maximum, and therefore when the orbital period is more or less nominal (i.e., equal to the mean orbital period quoted in Table~\ref{tab:paramsv471}), our measurements would have sampled intermediate values of the shear (expected to scale with the orbital period) rather than the maximum possible value for the K2~dwarf in the context of the Applegate framework. Surface shears larger by almost an order of magnitude  than those we detected are thus expected to be present when the orbital period is minimum if the Applegate mechanism is to explain the reported orbital period fluctuations, which remains to be investigated with more observations. 

An alternative mechanism requiring weak shear fluctuations ($\dO/\Omega \sim 0.004\%$) was recently suggested to operate in V471~Tau \citep{L20}. Like in the Applegate model, the new mechanism proposes that ETVs are caused by gravity changes in the orbital plane, with the main difference being the nature of the variations. \citet{L20} showed that if the K2 dwarf harbors a stationary non-axisymmetric magnetic field (instead of the dynamo modulated field invoked in the Applegate model), then an internal torque is introduced in the system forcing the magnetic structures to oscillate. Two distinct solutions were found for the magnetic structures whose orientation changes with respect to the star, a libration around phase 0.5 or circulation at a constant rate. In V471~Tau, \citet{L20} found a $70$-yr modulation period for the magnetic field for both libration and circulation models to account for the observed orbital period fluctuations of 35-yr. These models required magnetic field strengths at the base of the convective zone in the range of 8 to 17~kG, implying surface fields of a few kG. In order to asses whether this model is quantitatively compatible with observations, additional data similar to those analysed in this work must be acquired over the time span of the orbital period modulation.

\subsection{Magnetic activity and prominences}
\begin{figure}
	\centering
	\includegraphics[width=0.4\textwidth]{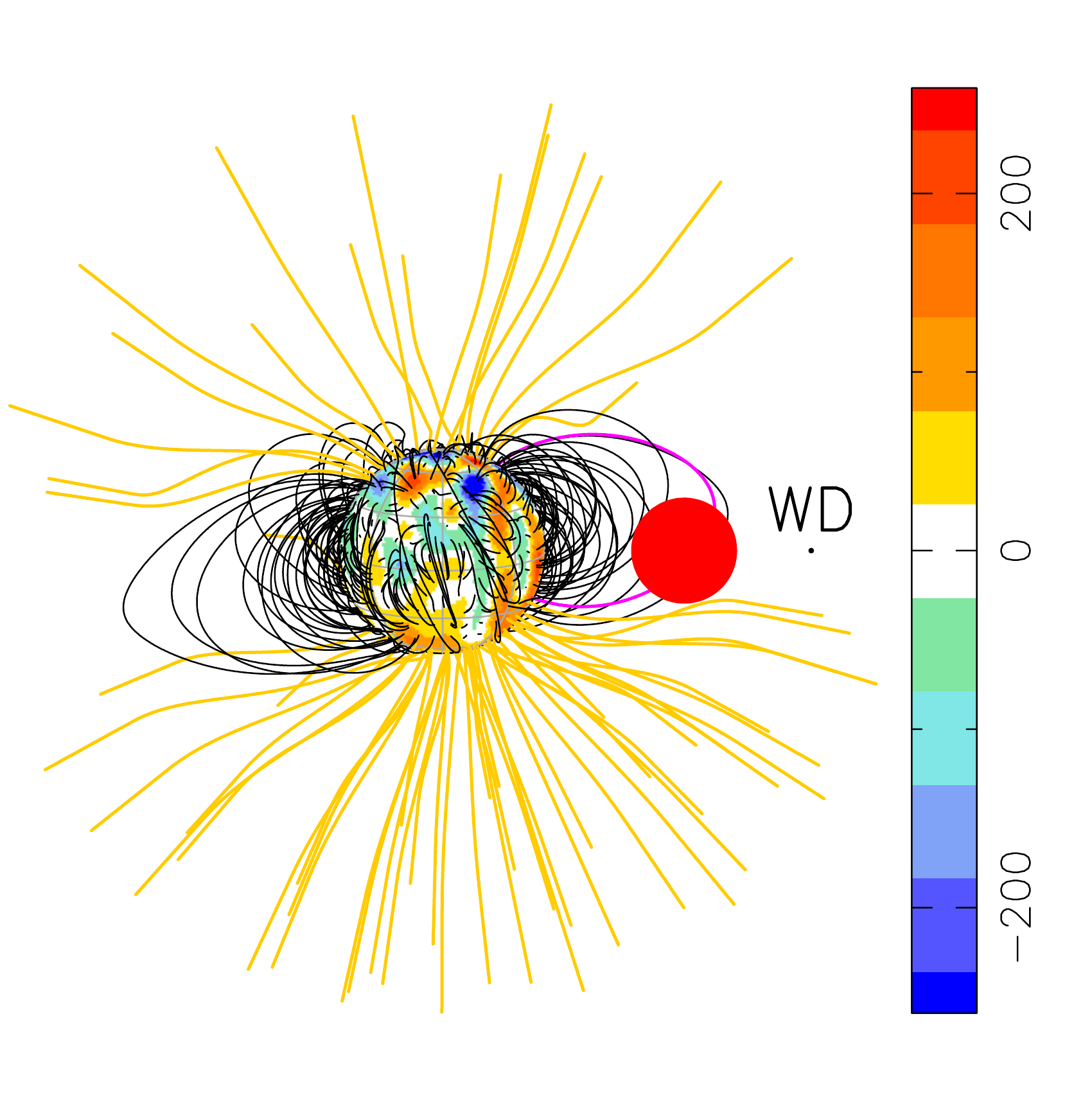}
	\includegraphics[width=0.4\textwidth]{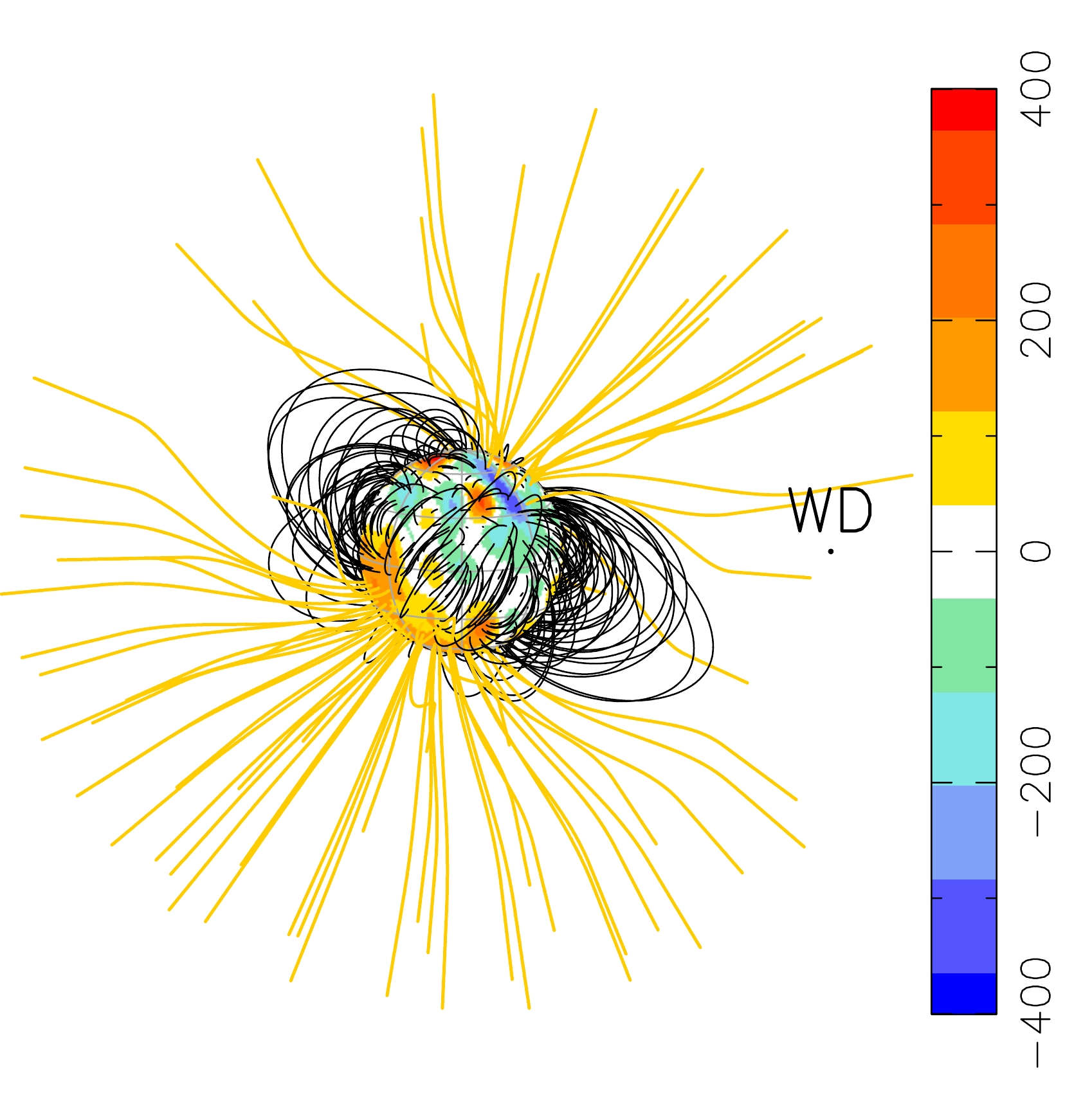} 
    \caption{Potential field extrapolations of the large scale radial field obtained with ZDI reconstructions in 2004 (top panel) and 2005 (bottom panel). Field lines are seen at rotational phase 0.75. The local strength of the magnetic field (G) at the surface of the star is shown in colours and open/closed lines are depicted in yellow/black. For this extrapolation, we assumed a source surface located 3.5~$R_\star$ beyond which all field lines break open, e.g., under the impact of centrifugal forces. The WD star (black circle) and the prominence detected in 2004 (red circle) are also shown. Field lines crossing the prominence are coloured in magenta.}
    \label{fig:maglines}
\end{figure}

We find that in our spectra of V471~Tau, H$\alpha$ exhibits a behaviour similar to that reported in the literature \citep[e.g.,][]{YRS91,BLM91,VWR02,KRJ07}, i.e. strongest when the WD is in front of the K2 star. \citet{RBY02} suggested that tidal forces in the binary system are able to trigger active longitudes at the surface of the K2 dwarf where the activity is enhanced with respect to the other side of the star. Potential field extrapolation of the surface radial field can help us visualize the magnetic field topologies obtained in our study (Fig.~\ref{fig:maglines}). Indeed, in 2005, the dipole field component, which largely dictates the overall geometry of the corona at a distance of a few stellar radii, seems to be oriented towards the azimuth of the WD. However, in 2004, nothing obvious shows up from the distribution of field lines. Admittedly, these potential field extrapolations are likely to be no more than rough descriptions of the magnetosphere, since we did not take in to account the gravitational impact of the WD.

\begin{figure}
	\centering
	\includegraphics[width=0.45\textwidth]{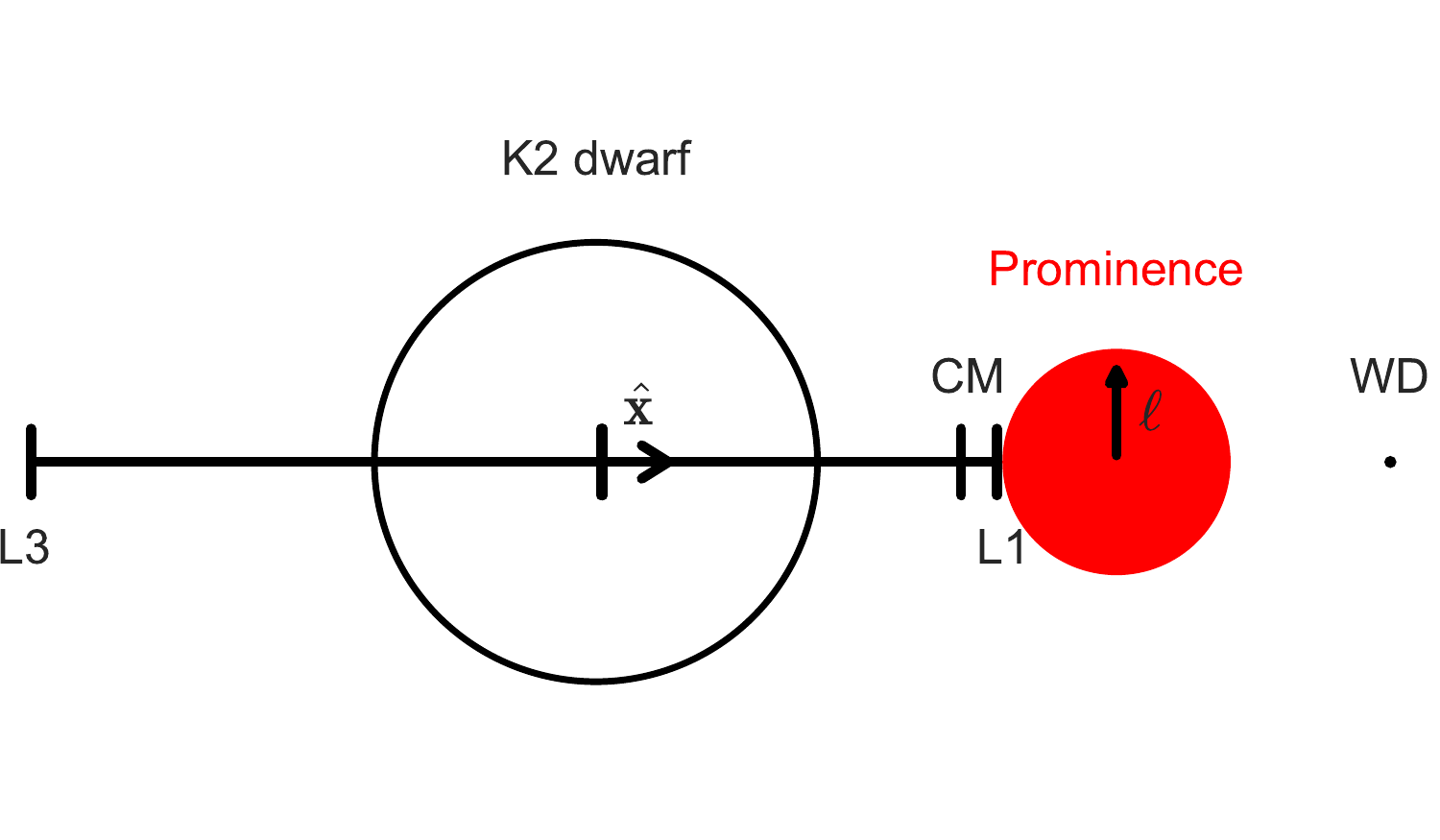} 
    \caption{Schematic view of the binary system V471~Tau in the presence of a prominence. Distances to the center of mass  and to the Lagrange points $L_1$ and $L_3$ are indicated by CM, L1, and L3 ($=-2.58~\mathrm{R_\star}~\vu{x}$), respectively. The size of the binaries and the prominence were kept in scale, where the prominence was approximated by a spherical blob with radius $\ell \sim 0.50 R_\star$ (see text).}
    \label{fig:schematic}
\end{figure}

Figure~\ref{fig:schematic} shows the schematic view of the system. In our Nov/Dec 2004 observations, the H$\alpha$ dynamical spectrum reveals the presence of a prominence at a stable location in the rotation frame over the seven rotation cycles of our observing run. Similar results are reported in previous studies on the activity of V471~Tau \citep[e.g.,][]{YRS91,RBY02}. We find the prominence to be located at $2.35 \pm 0.43~\mathrm{R_\star}$ from the centre of the K2 dwarf, farther away towards the WD than the center of mass of the system (located at $1.679\pm0.004~\mathrm{R_\star}$ from the centre of the K2 dwarf) and the Lagrange point L1 (located at $1.84\pm0.02~\mathrm{R_\star}$). Therefore, closed loops of the stellar magnetosphere likely extend out to few stellar radii and can maintain the slingshot prominence for at least 7 rotation cycles \citep{SHM96,JCD20}. Indeed, we can identify from the potential field reconstruction in 2004 some closed field lines that reach, and are potentially able to confine, the observed prominence (see Fig.~\ref{fig:maglines}). 

The spectral signature of the prominence indicates a Balmer decrement $I(\mathrm{H}\alpha)/I(\mathrm{H}\beta) \approx 3.4$. Although the assumption of optically thin emission is not strictly true given the Balmer decrement we measure, we nonetheless use it to derive a lower estimate for the mass density of the detected prominence, whose emission is mostly due to radiative recombination of hydrogen atoms. We assume that the prominence is a spherical blob with radius $l = 0.50~R_{\star}$ (computed from the FWHM of H$\alpha$) and a temperature of 5,000~K. From the prominence flux $f_\mathrm{H\alpha} = 2.1 \times 10^{-13}~\mathrm{erg}~\mathrm{s}^{-1}\mathrm{cm}^{-2}$, we use the mass density expression derived by \citet{SHM96} (their Eq.~3) to compute $\rho = 4\times10^{-14}~\mathrm{g}~\mathrm{cm^{-3}}$. Accordingly, we can use the volume of the blob to derive an estimate of the prominence mass of $6\times 10^{18}$~g. Our mass estimation is three to four orders of magnitude greater than large prominences in the Sun \citep[$\approx 10^{15}$~g,][]{HBB06} and one order of magnitude grater than the mass range derived for the prominences around AB~Dor \citep[2-10$\times 10^{17}$~g, e.g., ][]{CR89,CDE90}.

We estimate the minimum magnetic tension in the loop necessary to hold the prominence through the inequality
\begin{equation} \label{eq:tension}
    \frac{B^2}{4\pi R_c} \geq \rho g_\mathrm{eff},
\end{equation}
where $B$ and $R_\mathrm{c}$ are respectively the field strength and the curvature radius at the top of the loop, $\rho$ is the prominence density, and $g_\mathrm{eff}$ is the effective gravity acting on the gas accounting also for the centrifugal acceleration \citep[see][]{SHM96}. We follow previous authors and adopt a typical field curvature radius of $R_\mathrm{c} = 0.30~\mathrm{R_\star}$  \citep[e.g.,][]{DMC00}, keeping in mind that if the loop is further bended a lower field strength would be required to hold the prominence. The effective gravity inside the prominence features a sharp gradient, ranging from $\va{g_\mathrm{eff}} = 2.5$~m/s$^2~\vu{x}$ in the prominence regions closest to the L1 point up to $\va{g_\mathrm{eff}} = 485$~m/s$^2~\vu{x}$ in the prominence regions facing the WD, with an intermediate value of $\va{g_\mathrm{eff}} = 132$~m/s$^2~\vu{x}$ in the prominence central regions. With $\rho = 4 \times 10^{-14}~\mathrm{g}~\mathrm{cm}^{-3}$, we find that the $B$ field requested to hold the prominence material ranges from 2 to 22~G
depending on whether the magnetic loop crosses the orbital plane on the sides of the prominence that are closest to or farthest from the K2 star, and to a strength of 11~G at the center of the prominence. These field estimates tend to be larger than the values of the extrapolated field (see Fig.~\ref{fig:maglines}) we derive within the prominence (ranging from 8 to 2~G for the sides of the prominence that are closest to or farthest from the K2 star, respectively), especially in the prominence section closest to the WD. This suggests that the prominence region closest to the L1 point is the most stable against centrifugal ejection, unless the field lines at the top of the loop are bent to a larger extent than what we assumed here. Our results therefore suggest that the observed prominence is indeed likely sustained against centrifugal ejection by a slingshot mechanism, as already documented for several CVs in the past \citep[][]{SHM96}.

As the WD is known to be magnetic we can expect magnetosphere from both system stars to interact. Previous studies explored possible signatures from reconnection events at the magnetosphere interface \citep{PCS93,LWC96,NS99}. In the light of our result and given the dipole field of 350~kG reported for the WD \citep{SSB98,SBL12}, we can estimate the location of the magnetosphere interface between both components of V471~Tau, which we find to be located at a distance of about $3.07~\mathrm{R_\star}$ from the K2 star and $0.516~\mathrm{R_\star}$ from the WD (given the dipole field of $\sim\negmedspace100$~G found on the K2 star). From this magnetospheric interaction, we can expect increased activity at the surface of the K2 dwarf when the WD is in front of the K2 star (phase 0.5), in qualitative agreement with what is observed (see Fig.~\ref{fig:haeq}).

\section{Summary}
In our work, we reconstructed for the first time the large-scale magnetic field at the surface of the K2~dwarf and reported new detection of its surface differential rotation. The strong shear we found demonstrates that the K2~dwarf can be differentially rotating and is not always rotating as a solid body, as previous observations may have suggested \citep{HAS06}. Moreover, our results provide evidence for temporal variations in the surface shear of the K2~dwarf, making it even more similar to its single-star analog AB~Dor. Our findings highlight the importance of further investigation of the V471~Tau system in order to understand the impact of binary companions on the dynamo processes operating in the convective layers of low-mass stars. In particular, as differential rotation is expected to amplify toroidal magnetic fields, new spectropolarimetric observations of V471 Tau should allow us to study a potential connection between temporal variations in differential rotation and the underlying dynamo processes (e.g., through observations of polarity changes in the magnetic topology).

Furthermore, additional magnetic maps will probe the long-term effects of the K2~dwarf's magnetism on the V471~Tau system, invoked as the potential cause of the observed ETVs \citep[e.g.,][]{A92,L20}. As a follow-up study, we plan to monitor the system at a number of different phases of the ETV modulation cycle, in particular those at which the orbital period gets close to its minimum and maximum values, expected to correspond to the phase of the maximum and minimum shear in the framework of the Applegate mechanism. Monitoring the temporal evolution of the large-scale magnetic topology of the K2 dwarf of V471~Tau will also allow us to improve our understanding of the prominence stability and lifetime and determine their impact on the overall rate at which such prominences participate in the angular momentum loss of the whole system.

\section*{Acknowledgements}
We acknowledge funding from the European Research Council (ERC) under the H2020 research $\&$ innovation  programme (grant agreement $\#740651$ New-Worlds). 

%%%%%%%%%%%%%%%%%%%%%%%%%%%%%%%%%%%%%%%%%%%%%%%%%%
\section*{Data Availability}
This paper includes data collected by the ESPaDOnS spectropolarimeter,
which is publicly available from the Canadian Astronomy Data Center (program IDs: 05BF14 $\&$ 04BD50).



\bibliographystyle{mnras}
\bibliography{ref} % if your bibtex file is called example.bib

%\begin{thebibliography}{99}
%\end{thebibliography}

%%%%%%%%%%%%%%%%%%%%%%%%%%%%%%%%%%%%%%%%%%%%%%%%%%

%%%%%%%%%%%%%%%%% APPENDICES %%%%%%%%%%%%%%%%%%%%%
\appendix
\section{Journal of observations} \label{sec:tables}

This appendix contains the information on the V471~Tau spectropolarimetric data used in this study. Table~\ref{table:data1} shows the summary of observations for the Nov/Dec 2004 campaign and Table~\ref{table:data2} shows the information for the observational campaign in Dec 2005.

\begin{table*}	
	\caption{Summary of ESPaDOnS/CFHT observations for V471~Tau in 2004. Columns 1 to 4 respectively record  (i) the date of observation, (ii) the UT time at mid sub-exposure, (iii) the time in Heliocentric Julian Date (HJD) in excess of  $2,453,337$~days, and (iv) the rotation cycle of each observation (computed using Eq.~\ref{eq:ephemeris}). Total exposure times ($t_\mathrm{exp}$) correspond to the sum of the four sub-exposures times used to compute each circularly polarisation profile. Column 5 illustrates peak SNR values for the Stokes $V$ spectrum (per $1.8$~km/s spectral pixel) and column 6 indicates the profiles excluded in the ZDI analysis. The root-mean-square (rms) noise level in the circular polarisation profiles produced by Least-Squares Deconvolution (LSD) is given in column 8. The last two columns list the longitudinal field $B_\ell$ and its corresponding 1$\sigma$ error bar.}
	\begin{tabular}{lccccccccr} 
		\hline
		Date & UT        & HJD    & $E$        &  $t_\mathrm{exp}$  &    SNR &  Comment & $\sigma_{LSD}$ & $B_\ell$ & $\sigma_{B_\ell}$ \\
		(2004) & (h:m:s) & $(2,453,337+) $ & &   (s)     &   & & $(10^{-4}) $&   (G) & (G)      \\
		\hline
Nov 28   &  08:53:48 &  0.87674  & 0.986220 & 4 $\times$ 200 & 180 &            & 1.6  &  -40 &   29  \\
Nov 28   &  09:16:41 &  0.89262  & 1.016689 & 4 $\times$ 200 & 178 &            & 1.6  &    5 &   30  \\
Nov 28   &  09:34:47 &  0.90520  & 1.040826 & 4 $\times$ 200 & 185 &            & 1.6  &    5 &   28  \\
Nov 29   &  05:54:13 &  1.75200  & 2.665590 & 4 $\times$ 200 & 172 &            & 1.8  &   25 &   32  \\
Nov 29   &  06:11:40 &  1.76412  & 2.688845 & 4 $\times$ 200 & 172 &            & 1.8  &   -9 &   31  \\
Nov 29   &  06:29:11 &  1.77627  & 2.712157 & 4 $\times$ 200 & 178 &            & 1.7  &  -24 &   30  \\
Nov 29   &  06:46:46 &  1.78849  & 2.735604 & 4 $\times$ 200 & 182 &            & 1.7  &   24 &   30  \\
Nov 29   &  07:07:13 &  1.80269  & 2.762849 & 4 $\times$ 200 & 185 &            & 1.6  &    7 &   28  \\
Nov 29   &  07:24:39 &  1.81479  & 2.786066 & 4 $\times$ 200 & 180 &            & 1.6  &   15 &   28  \\
Nov 29   &  07:42:06 &  1.82691  & 2.809321 & 4 $\times$ 200 & 191 &            & 1.5  &  -38 &   27  \\
Nov 29   &  07:59:31 &  1.83901  & 2.832537 & 4 $\times$ 200 & 182 &            & 1.6  &   -3 &   28  \\
Nov 29   &  08:16:56 &  1.85110  & 2.855734 & 4 $\times$ 200 & 181 &            & 1.6  &    0 &   28  \\
Nov 29   &  08:59:03 &  1.88035  & 2.911857 & 4 $\times$ 200 & 173 &            & 1.7  &  -82 &   30  \\
Nov 29   &  09:16:28 &  1.89244  & 2.935054 & 4 $\times$ 200 & 181 &            & 1.6  &  -60 &   28  \\
Nov 29   &  09:33:52 &  1.90453  & 2.958251 & 4 $\times$ 200 & 165 &            & 1.8  &  -46 &   32  \\
Nov 29   &  10:05:09 &  1.92625  & 2.999925 & 4 $\times$ 200 & 148 &            & 2.1  &   18 &   38  \\
Nov 29   &  10:24:18 &  1.93954  & 3.025425 & 4 $\times$ 200 & 103 &            & 3.0  &  -92 &   55  \\
Nov 29   &  10:41:50 &  1.95172  & 3.048795 & 4 $\times$ 200 & 113 &            & 2.7  &   31 &   49  \\
Nov 29   &  10:59:22 &  1.96390  & 3.072165 & 4 $\times$ 200 & 127 &            & 2.3  &    0 &   42  \\
Nov 29   &  11:21:27 &  1.97923  & 3.101579 & 4 $\times$ 200 &  48 &  Bad SNR   &      &      &       \\
Nov 29   &  11:38:58 &  1.99139  & 3.124910 & 4 $\times$ 200 &  75 &            & 4.3  &   34 &   78  \\
Nov 29   &  11:56:27 &  2.00354  & 3.148222 & 4 $\times$ 200 &  95 &            & 3.4  &  115 &   61  \\
Nov 29   &  12:14:41 &  2.01620  & 3.172513 & 4 $\times$ 200 &  90 &            & 3.6  &  -48 &   65  \\
Nov 29   &  12:32:12 &  2.02837  & 3.195864 & 4 $\times$ 200 &  39 &  Bad SNR   &      &      &       \\
Nov 29   &  12:49:45 &  2.04056  & 3.219253 & 4 $\times$ 200 &  75 &            & 4.9  &  -76 &   89  \\
Nov 29   &  13:42:45 &  2.07736  & 3.289862 & 4 $\times$ 200 &  72 &  Bad SNR   &      &      &       \\
Nov 29   &  14:00:22 &  2.08959  & 3.313327 & 4 $\times$ 200 & 117 &            & 2.6  &   20 &   48  \\
Nov 29   &  14:17:56 &  2.10178  & 3.336717 & 4 $\times$ 200 & 106 &            & 3.2  &  -93 &   58  \\
Dec 01   &  06:13:29 &  3.76531  & 6.528549 & 4 $\times$ 200 & 134 &            & 2.4  &  -32 &   44  \\
Dec 01   &  06:31:03 &  3.77751  & 6.551957 & 4 $\times$ 200 & 163 &            & 1.9  &   17 &   34  \\
Dec 01   &  06:48:43 &  3.78978  & 6.575500 & 4 $\times$ 120 & 113 &            & 2.9  &   66 &   53  \\
Dec 01   &  07:00:50 &  3.79820  & 6.591655 & 4 $\times$ 120 & 109 &            & 2.9  &  -78 &   52  \\
Dec 01   &  07:12:58 &  3.80662  & 6.607811 & 4 $\times$ 120 &  85 &            & 3.8  &   33 &   69  \\
Dec 01   &  07:25:07 &  3.81505  & 6.623986 & 4 $\times$ 120 &  69 &  Bad SNR   &      &      &       \\
Dec 01   &  07:37:14 &  3.82347  & 6.640141 & 4 $\times$ 120 & 100 &            & 3.2  &  -20 &   57  \\
Dec 01   &  07:51:09 &  3.83313  & 6.658676 & 4 $\times$ 120 & 138 &            & 2.1  &   35 &   38  \\
Dec 01   &  08:03:16 &  3.84154  & 6.674812 & 4 $\times$ 120 & 143 &            & 2.1  &   21 &   38  \\
Dec 01   &  08:39:43 &  3.86686  & 6.723394 & 4 $\times$ 200 & 164 &            & 1.8  &  -55 &   32  \\
Dec 01   &  08:57:12 &  3.87900  & 6.746687 & 4 $\times$ 200 & 127 &            & 2.5  &  -32 &   45  \\
Dec 01   &  09:14:40 &  3.89113  & 6.769961 & 4 $\times$ 200 & 150 &            & 2.0  &   40 &   36  \\
Dec 01   &  09:32:09 &  3.90326  & 6.793235 & 4 $\times$ 200 & 143 &            & 2.1  &  -52 &   37  \\
Dec 01   &  09:49:36 &  3.91539  & 6.816509 & 4 $\times$ 200 & 133 &            & 2.3  &   -4 &   41  \\
Dec 01   &  10:07:36 &  3.92789  & 6.840493 & 4 $\times$ 200 &  76 &            & 4.4  &  -63 &   80  \\
Dec 01   &  10:25:03 &  3.94000  & 6.863728 & 4 $\times$ 200 &  91 &            & 3.8  &   27 &   68  \\
Dec 01   &  10:42:27 &  3.95209  & 6.886926 & 4 $\times$ 200 & 173 &            & 1.7  &  -41 &   30  \\
Dec 01   &  10:59:52 &  3.96418  & 6.910123 & 4 $\times$ 200 & 165 &            & 1.8  &  -81 &   32  \\
Dec 01   &  11:17:19 &  3.97629  & 6.933358 & 4 $\times$ 200 & 164 &            & 1.8  &  -74 &   32  \\
Dec 01   &  11:35:34 &  3.98897  & 6.957688 & 4 $\times$ 200 & 166 &            & 1.7  &  -59 &   31  \\
Dec 01   &  11:53:00 &  4.00108  & 6.980923 & 4 $\times$ 200 & 157 &            & 1.8  &    2 &   33  \\
Dec 01   &  12:10:27 &  4.01320  & 7.004178 & 4 $\times$ 200 & 147 &            & 2.0  &  -32 &   36  \\
Dec 01   &  12:27:53 &  4.02531  & 7.027414 & 4 $\times$ 200 & 146 &            & 1.9  &    3 &   35  \\
Dec 01   &  12:45:18 &  4.03740  & 7.050611 & 4 $\times$ 200 & 136 &            & 2.0  &   34 &   36  \\
Dec 01   &  13:06:35 &  4.05217  & 7.078950 & 4 $\times$ 200 & 135 &            & 2.4  &  -10 &   43  \\
Dec 01   &  13:24:11 &  4.06440  & 7.102416 & 4 $\times$ 200 & 149 &            & 2.0  &   -1 &   36  \\
Dec 01   &  13:41:52 &  4.07668  & 7.125978 & 4 $\times$ 200 & 146 &            & 2.0  &   37 &   36  \\
Dec 01   &  14:09:47 &  4.09607  & 7.163182 & 4 $\times$ 200 & 139 &            & 2.2  &  -50 &   39  \\
		\hline		
	\end{tabular}
	\label{table:data1}
\end{table*}

\begin{table*}	
	\caption{Same as Table~\ref{table:data1} but for acquisitions in 2005.}
	\begin{tabular}{lccccccccr} % four columns, alignment for each
		\hline
		Date & UT        & HJD    & $E$        &  $t_\mathrm{exp}$  &    SNR &  Comment & $\sigma_{LSD}$ & $B_\ell$ & $\sigma_{B_\ell}$  \\
		(2005) & (h:m:s) & $(2,453,337+) $ & &   (s)     &   & & $(10^{-4}) $&   (G) & (G)      \\
		\hline
		
Dec 14   &  05:16:29 &  381.72517  &  731.724073  & 4 $\times$ 200  & 138 &          & 2.3  &    12 &   41  \\
Dec 14   &  05:35:60 &  381.73872  &  731.750072  & 4 $\times$ 200  & 142 &          & 2.2  &    60 &   39  \\
Dec 14   &  05:52:50 &  381.75041  &  731.772502  & 4 $\times$ 200  & 146 &          & 2.1  &    49 &   38  \\
Dec 14   &  06:09:39 &  381.76209  &  731.794912  & 4 $\times$ 200  & 144 &          & 2.1  &    28 &   38  \\
Dec 14   &  06:26:27 &  381.77375  &  731.817284  & 4 $\times$ 200  & 141 &          & 2.2  &    71 &   39  \\
Dec 14   &  06:46:03 &  381.78737  &  731.843417  & 4 $\times$ 200  & 107 &          & 3.1  &    88 &   56  \\
Dec 14   &  07:02:52 &  381.79904  &  731.865808  & 4 $\times$ 200  & 122 &          & 2.7  &   131 &   48  \\
Dec 14   &  07:19:39 &  381.81070  &  731.888181  & 4 $\times$ 200  & 145 &          & 2.1  &    40 &   38  \\
Dec 14   &  07:36:27 &  381.82236  &  731.910553  & 4 $\times$ 200  & 140 &          & 2.2  &     5 &   40  \\
Dec 14   &  07:55:41 &  381.83572  &  731.936187  & 4 $\times$ 200  & 139 &          & 2.3  &     1 &   40  \\
Dec 14   &  08:30:56 &  381.86019  &  731.983138  & 4 $\times$ 200  & 141 &          & 2.3  &    -8 &   40  \\
Dec 14   &  08:47:52 &  381.87196  &  732.005721  & 4 $\times$ 200  & 131 &          & 2.5  &    14 &   45  \\
Dec 14   &  09:04:40 &  381.88362  &  732.028093  & 4 $\times$ 200  & 138 &          & 2.3  &    11 &   41  \\
Dec 14   &  09:21:37 &  381.89539  &  732.050676  & 4 $\times$ 200  & 122 &          & 2.5  &     5 &   46  \\
Dec 14   &  09:49:16 &  381.91459  &  732.087515  & 4 $\times$ 200  & 131 &          & 2.4  &   -69 &   43  \\
Dec 14   &  10:06:11 &  381.92634  &  732.110060  & 4 $\times$ 200  & 134 &          & 2.3  &   -69 &   42  \\
Dec 14   &  10:23:14 &  381.93817  &  732.132759  & 4 $\times$ 200  & 123 &          & 2.6  &    48 &   47  \\
Dec 14   &  10:40:13 &  381.94997  &  732.155399  & 4 $\times$ 200  & 128 &          & 2.6  &  -101 &   46  \\
Dec 14   &  10:57:48 &  381.96218  &  732.178827  & 4 $\times$ 200  & 113 &          & 2.8  &   -42 &   51  \\
Dec 14   &  11:14:36 &  381.97385  &  732.201218  & 4 $\times$ 200  & 105 &          & 3.1  &   -12 &   56  \\
Dec 14   &  11:31:38 &  381.98567  &  732.223897  & 4 $\times$ 200  &  75 &          & 4.8  &  -140 &   87  \\
Dec 14   &  11:48:25 &  381.99734  &  732.246289  & 4 $\times$ 200  &  73 &  Bad SRN &      &       &       \\
Dec 14   &  12:06:04 &  382.00958  &  732.269774  & 4 $\times$ 200  & 104 &          & 3.3  &   -91 &   60  \\
Dec 14   &  12:22:54 &  382.02127  &  732.292203  & 4 $\times$ 200  & 104 &          & 3.2  &   -67 &   58  \\
Dec 14   &  12:40:28 &  382.03348  &  732.315631  & 4 $\times$ 200  &  73 &  Bad SNR &      &       &       \\
Dec 14   &  12:58:17 &  382.04585  &  732.339365  & 4 $\times$ 200  &  61 &  Bad SNR &      &       &       \\
Dec 16   &  04:45:48 &  383.70375  &  735.520395  & 4 $\times$ 200  & 140 &          & 2.3  &   -26 &   41  \\
Dec 16   &  05:02:23 &  383.71527  &  735.542499  & 4 $\times$ 200  & 161 &          & 1.9  &   -25 &   35  \\
Dec 16   &  05:19:00 &  383.72681  &  735.564641  & 4 $\times$ 200  & 167 &          & 1.8  &   -69 &   33  \\
Dec 16   &  05:35:39 &  383.73837  &  735.586821  & 4 $\times$ 200  & 172 &          & 1.8  &   -32 &   31  \\
Dec 16   &  05:55:51 &  383.75239  &  735.613721  & 4 $\times$ 200  & 173 &          & 1.8  &   -20 &   31  \\
Dec 16   &  06:12:37 &  383.76404  &  735.636074  & 4 $\times$ 200  & 174 &          & 1.7  &     9 &   30  \\
Dec 16   &  06:29:28 &  383.77574  &  735.658523  & 4 $\times$ 200  & 177 &          & 1.7  &    26 &   30  \\
Dec 16   &  06:46:11 &  383.78735  &  735.680800  & 4 $\times$ 200  & 179 &          & 1.7  &    21 &   29  \\
Dec 16   &  07:06:05 &  383.80116  &  735.707297  & 4 $\times$ 200  & 180 &          & 1.7  &    61 &   29  \\
Dec 16   &  07:22:42 &  383.81270  &  735.729439  & 4 $\times$ 200  & 184 &          & 1.6  &    14 &   29  \\
Dec 16   &  07:39:25 &  383.82431  &  735.751715  & 4 $\times$ 200  & 179 &          & 1.6  &   -17 &   29  \\
Dec 16   &  08:03:45 &  383.84121  &  735.784141  & 4 $\times$ 200  & 182 &          & 1.6  &    52 &   29  \\
Dec 16   &  08:20:29 &  383.85283  &  735.806437  & 4 $\times$ 200  & 177 &          & 1.7  &    -6 &   30  \\
Dec 16   &  08:37:06 &  383.86437  &  735.828579  & 4 $\times$ 200  & 166 &          & 1.8  &    28 &   32  \\
Dec 16   &  08:54:00 &  383.87610  &  735.851085  & 4 $\times$ 200  & 172 &          & 1.8  &   -14 &   32  \\
Dec 16   &  09:13:16 &  383.88949  &  735.876777  & 4 $\times$ 200  & 165 &          & 1.8  &    90 &   33  \\
Dec 16   &  09:30:07 &  383.90119  &  735.899226  & 4 $\times$ 200  & 160 &          & 1.9  &    -7 &   34  \\
Dec 16   &  09:46:45 &  383.91273  &  735.921368  & 4 $\times$ 200  & 154 &          & 2.0  &    34 &   35  \\
Dec 16   &  10:03:21 &  383.92427  &  735.943509  & 4 $\times$ 200  & 155 &          & 2.0  &    27 &   35  \\
Dec 16   &  10:20:51 &  383.93642  &  735.966822  & 4 $\times$ 200  & 151 &          & 2.0  &    80 &   36  \\
Dec 16   &  10:37:39 &  383.94808  &  735.989194  & 4 $\times$ 200  & 159 &          & 1.9  &    65 &   35  \\
Dec 16   &  10:54:16 &  383.95962  &  736.011336  & 4 $\times$ 200  & 147 &          & 2.1  &    30 &   38  \\
Dec 16   &  11:10:54 &  383.97117  &  736.033497  & 4 $\times$ 200  & 141 &          & 2.2  &     2 &   40  \\
Dec 16   &  11:28:23 &  383.98331  &  736.056790  & 4 $\times$ 200  & 136 &          & 2.2  &    16 &   40  \\
Dec 16   &  11:45:09 &  383.99495  &  736.079124  & 4 $\times$ 200  & 138 &          & 2.2  &     6 &   40  \\
Dec 16   &  12:01:47 &  384.00650  &  736.101285  & 4 $\times$ 200  & 133 &          & 2.4  &   -25 &   43  \\
Dec 16   &  12:18:27 &  384.01807  &  736.123484  & 4 $\times$ 200  & 124 &          & 2.6  &   -25 &   47  \\
Dec 16   &  12:45:17 &  384.03670  &  736.159230  & 4 $\times$ 200  & 126 &          & 2.6  &    -7 &   47  \\
Dec 16   &  13:02:10 &  384.04843  &  736.181737  & 4 $\times$ 200  & 122 &          & 2.7  &   -42 &   50  \\
Dec 18   &  04:31:44 &  385.69386  &  739.338840  & 4 $\times$ 200  & 143 &          & 2.2  &   -26 &   40  \\
Dec 18   &  04:48:25 &  385.70546  &  739.361097  & 4 $\times$ 200  & 151 &          & 2.0  &   -83 &   36  \\
Dec 18   &  05:07:34 &  385.71875  &  739.386597  & 4 $\times$ 200  & 158 &          & 1.9  &     5 &   35  \\
Dec 18   &  05:24:13 &  385.73032  &  739.408796  & 4 $\times$ 200  & 165 &          & 1.8  &   -52 &   33  \\
Dec 18   &  05:41:03 &  385.74201  &  739.431226  & 4 $\times$ 200  & 173 &          & 1.7  &   -42 &   31  \\
Dec 18   &  05:57:45 &  385.75360  &  739.453464  & 4 $\times$ 200  & 174 &          & 1.7  &   -22 &   31  \\
		\hline		
	\end{tabular}
	\label{table:data2}
\end{table*}

\begin{table*}
	\contcaption{}
		\begin{tabular}{lccccccccr} % four columns, alignment for each
		\hline
		Date & UT        & HJD    & $E$        &  $t_\mathrm{exp}$  &    SNR &  Comment & $\sigma_{LSD}$ & $B_\ell$ & $\sigma_{B_\ell}$  \\
		(2005) & (h:m:s) & $(2,453,337+) $ & &   (s)     &   & & $(10^{-4}) $&   (G) & (G)      \\
		\hline

Dec 18   &  06:15:15 &  385.76576  &  739.476795  & 4 $\times$ 200  & 171 &          & 1.8  &   -28 &   32  \\
Dec 18   &  06:31:60 &  385.77738  &  739.499091  & 4 $\times$ 200  & 170 &          & 1.8  &     5 &   32  \\
Dec 18   &  06:49:04 &  385.78923  &  739.521828  & 4 $\times$ 200  & 175 &          & 1.7  &   -32 &   31  \\
Dec 18   &  07:05:53 &  385.80091  &  739.544238  & 4 $\times$ 200  & 163 &          & 1.8  &   -26 &   33  \\
Dec 18   &  07:27:07 &  385.81566  &  739.572539  & 4 $\times$ 200  & 163 &          & 1.8  &   -50 &   32  \\
Dec 18   &  08:04:07 &  385.84135  &  739.621831  & 4 $\times$ 200  & 180 &          & 1.6  &     9 &   29  \\
Dec 18   &  08:20:48 &  385.85293  &  739.644049  & 4 $\times$ 200  & 178 &          & 1.6  &   -10 &   29  \\
Dec 18   &  08:37:23 &  385.86445  &  739.666153  & 4 $\times$ 200  & 179 &          & 1.6  &   -10 &   29  \\
Dec 18   &  08:54:12 &  385.87613  &  739.688564  & 4 $\times$ 200  & 185 &          & 1.6  &    12 &   29  \\
Dec 18   &  09:12:37 &  385.88891  &  739.713085  & 4 $\times$ 200  & 184 &          & 1.6  &    12 &   28  \\
Dec 18   &  09:29:21 &  385.90054  &  739.735399  & 4 $\times$ 200  & 188 &          & 1.6  &    69 &   28  \\
Dec 18   &  09:46:12 &  385.91224  &  739.757848  & 4 $\times$ 200  & 183 &          & 1.6  &     6 &   29  \\
Dec 18   &  10:02:51 &  385.92380  &  739.780028  & 4 $\times$ 200  & 179 &          & 1.7  &     1 &   30  \\
Dec 18   &  10:20:26 &  385.93600  &  739.803437  & 4 $\times$ 200  & 178 &          & 1.7  &     8 &   30  \\
Dec 18   &  10:37:09 &  385.94762  &  739.825732  & 4 $\times$ 200  & 178 &          & 1.7  &    92 &   30  \\
Dec 18   &  10:53:49 &  385.95919  &  739.847932  & 4 $\times$ 200  & 180 &          & 1.7  &   -20 &   30  \\
Dec 18   &  11:10:31 &  385.97078  &  739.870169  & 4 $\times$ 200  & 181 &          & 1.7  &   -14 &   30  \\
Dec 18   &  11:28:13 &  385.98307  &  739.893750  & 4 $\times$ 200  & 181 &          & 1.7  &    20 &   30  \\
Dec 18   &  11:44:54 &  385.99466  &  739.915988  & 4 $\times$ 200  & 179 &          & 1.7  &    58 &   29  \\
Dec 18   &  12:01:42 &  386.00632  &  739.938360  & 4 $\times$ 200  & 179 &          & 1.7  &    -4 &   30  \\
Dec 18   &  12:18:29 &  386.01798  &  739.960733  & 4 $\times$ 200  & 175 &          & 1.7  &     0 &   30  \\
Dec 18   &  12:35:59 &  386.03014  &  739.984064  & 4 $\times$ 200  & 163 &          & 1.9  &     6 &   33  \\
Dec 20   &  04:29:28 &  387.69218  &  743.173038  & 4 $\times$ 200  & 150 &          & 2.1  &   -36 &   38  \\
Dec 20   &  04:47:10 &  387.70447  &  743.196619  & 4 $\times$ 200  & 150 &          & 2.1  &  -104 &   37  \\
Dec 20   &  05:03:45 &  387.71598  &  743.218703  & 4 $\times$ 200  & 156 &          & 2.0  &  -113 &   36  \\
Dec 20   &  05:20:31 &  387.72763  &  743.241056  & 4 $\times$ 200  & 164 &          & 1.9  &  -172 &   34  \\
Dec 20   &  05:37:32 &  387.73944  &  743.263716  & 4 $\times$ 200  & 162 &          & 1.9  &   -81 &   34  \\
Dec 20   &  05:55:18 &  387.75178  &  743.287393  & 4 $\times$ 200  & 157 &          & 2.0  &   -60 &   35  \\
Dec 20   &  06:11:58 &  387.76335  &  743.309592  & 4 $\times$ 200  & 154 &          & 2.0  &  -115 &   36  \\
Dec 20   &  06:28:41 &  387.77495  &  743.331849  & 4 $\times$ 200  & 144 &          & 2.1  &   -42 &   39  \\
Dec 20   &  06:45:27 &  387.78660  &  743.354202  & 4 $\times$ 200  & 155 &          & 2.0  &    -1 &   36  \\
Dec 20   &  07:03:14 &  387.79895  &  743.377898  & 4 $\times$ 200  & 158 &          & 1.9  &   -56 &   34  \\
Dec 20   &  07:43:06 &  387.82664  &  743.431027  & 4 $\times$ 200  & 161 &          & 1.9  &   -54 &   34  \\
Dec 20   &  08:00:32 &  387.83874  &  743.454244  & 4 $\times$ 200  & 159 &          & 1.9  &   -18 &   34  \\
Dec 20   &  08:18:17 &  387.85107  &  743.477902  & 4 $\times$ 200  & 153 &          & 2.0  &    24 &   36  \\
Dec 20   &  08:35:51 &  387.86326  &  743.501291  & 4 $\times$ 200  & 172 &          & 1.8  &   -52 &   32  \\
Dec 20   &  08:53:22 &  387.87543  &  743.524641  & 4 $\times$ 200  & 157 &          & 1.9  &   -18 &   34  \\		
		\hline		
	\end{tabular}
\end{table*}

\section{Fit to Stokes profiles} \label{sec:StokesAppendix}
\begin{figure*}
    \centering
    \includegraphics[width=1.\textwidth]{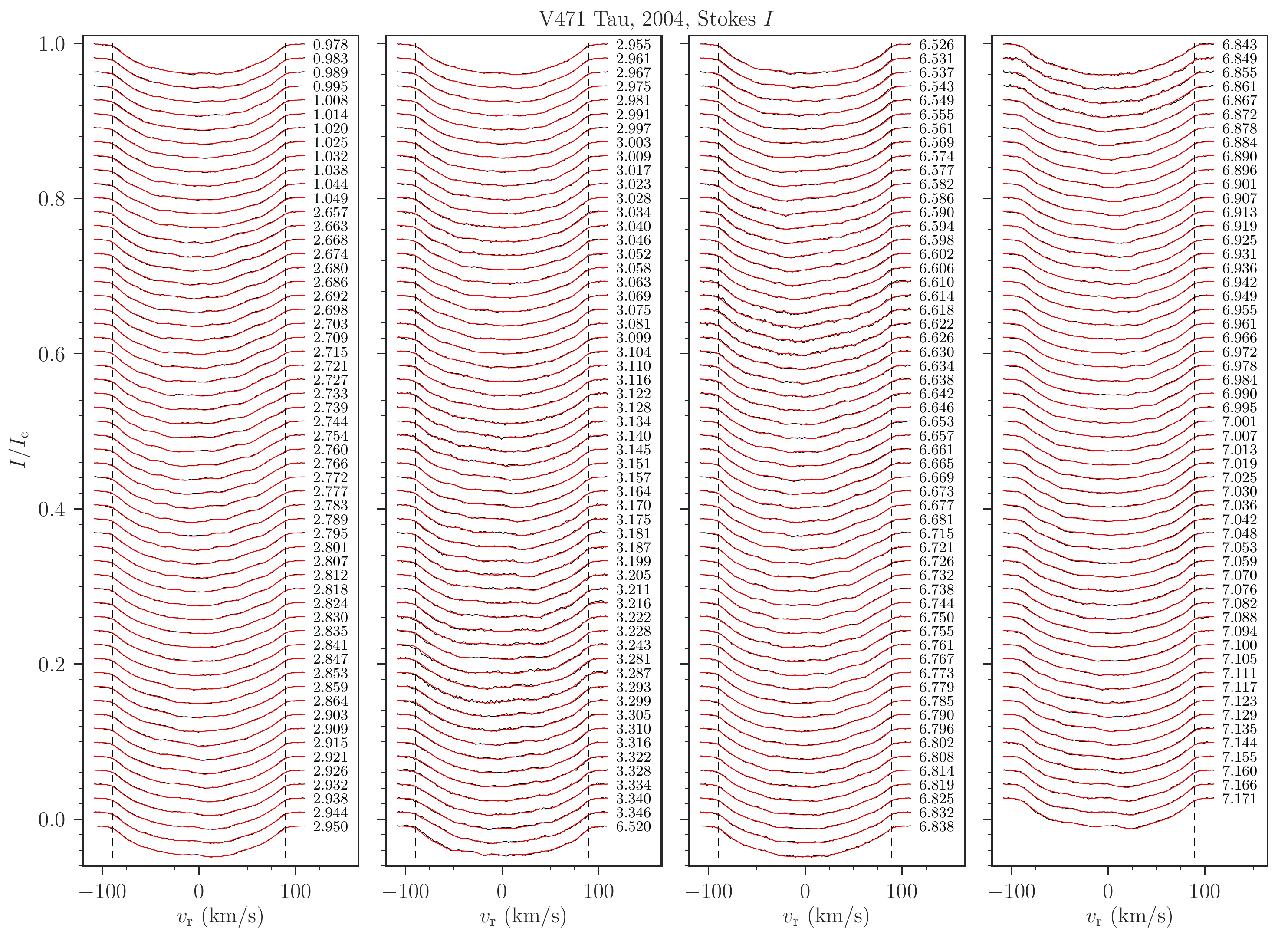} 
    \caption{Observed (black line) and modeled (red line) Stokes $I$ profiles of the V471 Tau K2 dwarf component, collected in Nov/Dec 2004. Individual observations are shifted vertically for display purposes. The rotation cycle of each observation is indicated on the right and the velocities of $\pm v\sin i$ are illustrated as dashed vertical lines.}
    \label{fig:si2004}
\end{figure*}

\begin{figure*}
    \includegraphics[width=1.\textwidth]{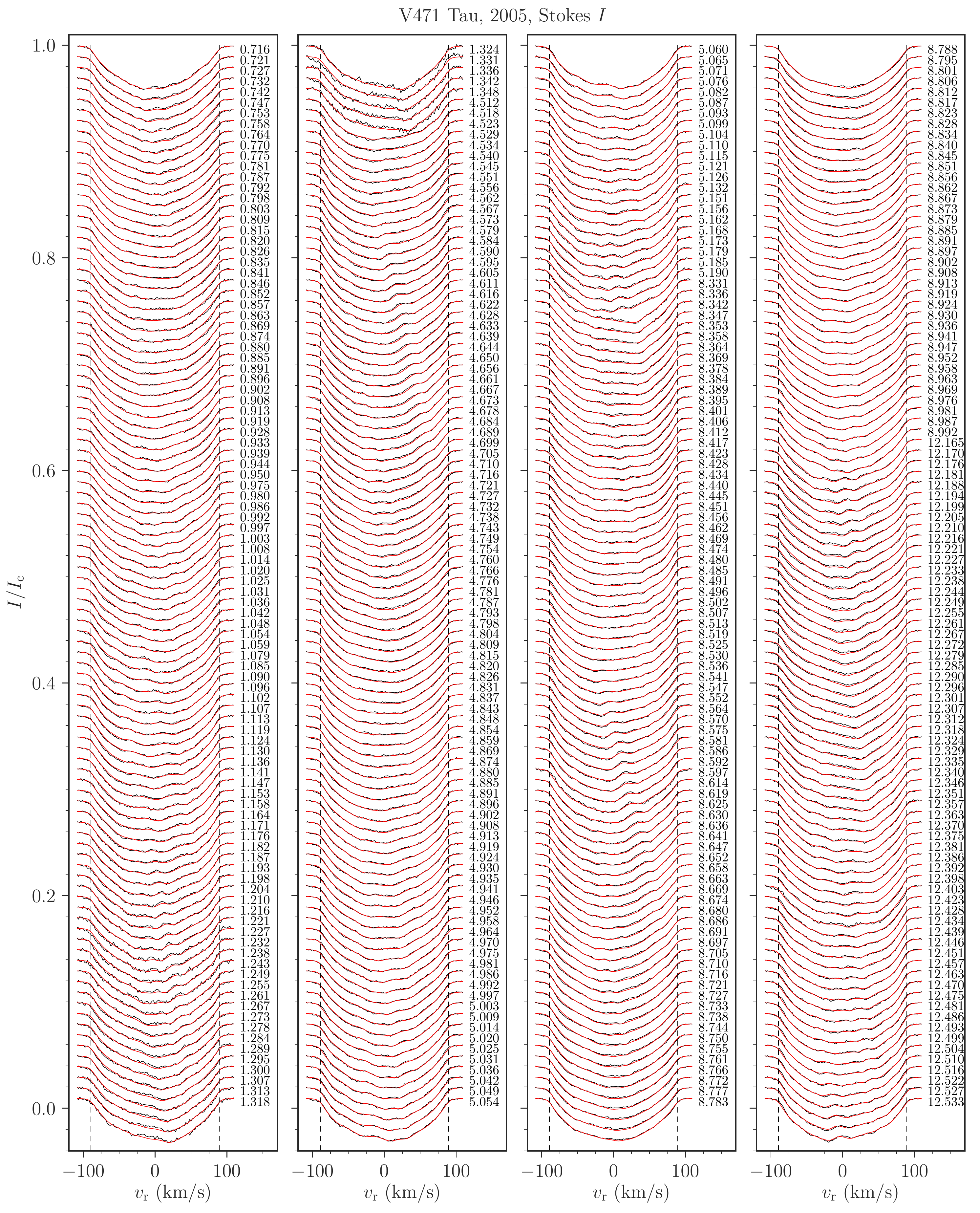}
    \caption{Same as Fig.~\ref{fig:si2004} but for the $400$ 
    Stokes $I$ spectra collected in Dec 2005.}
    \label{fig:si2005}
\end{figure*}

\begin{figure*}
    \centering
    \includegraphics[width=1.\textwidth]{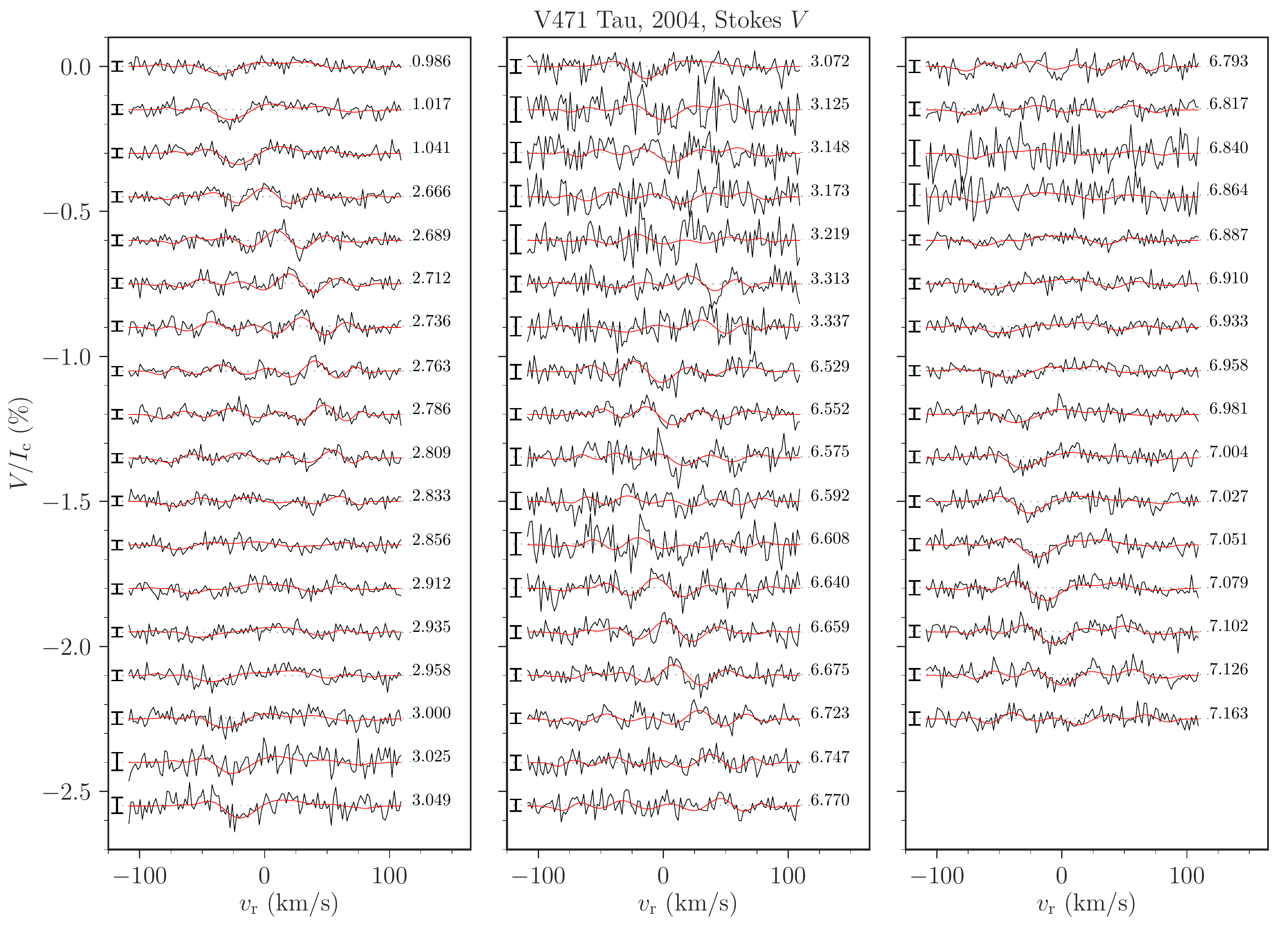} 
    \caption{Observed (black line) and modeled (red line) Stokes $V$ profiles of the V471 Tau K2 dwarf component, collected in Nov/Dec 2004. Individual observations are shifted vertically for display purposes. The rotation cycle of each observation is indicated on the right and $\pm$1\,$\sigma$ error-bars on the left of each profile.}
    \label{fig:sv2004}
\end{figure*}

\begin{figure*}
    \includegraphics[width=1.\textwidth]{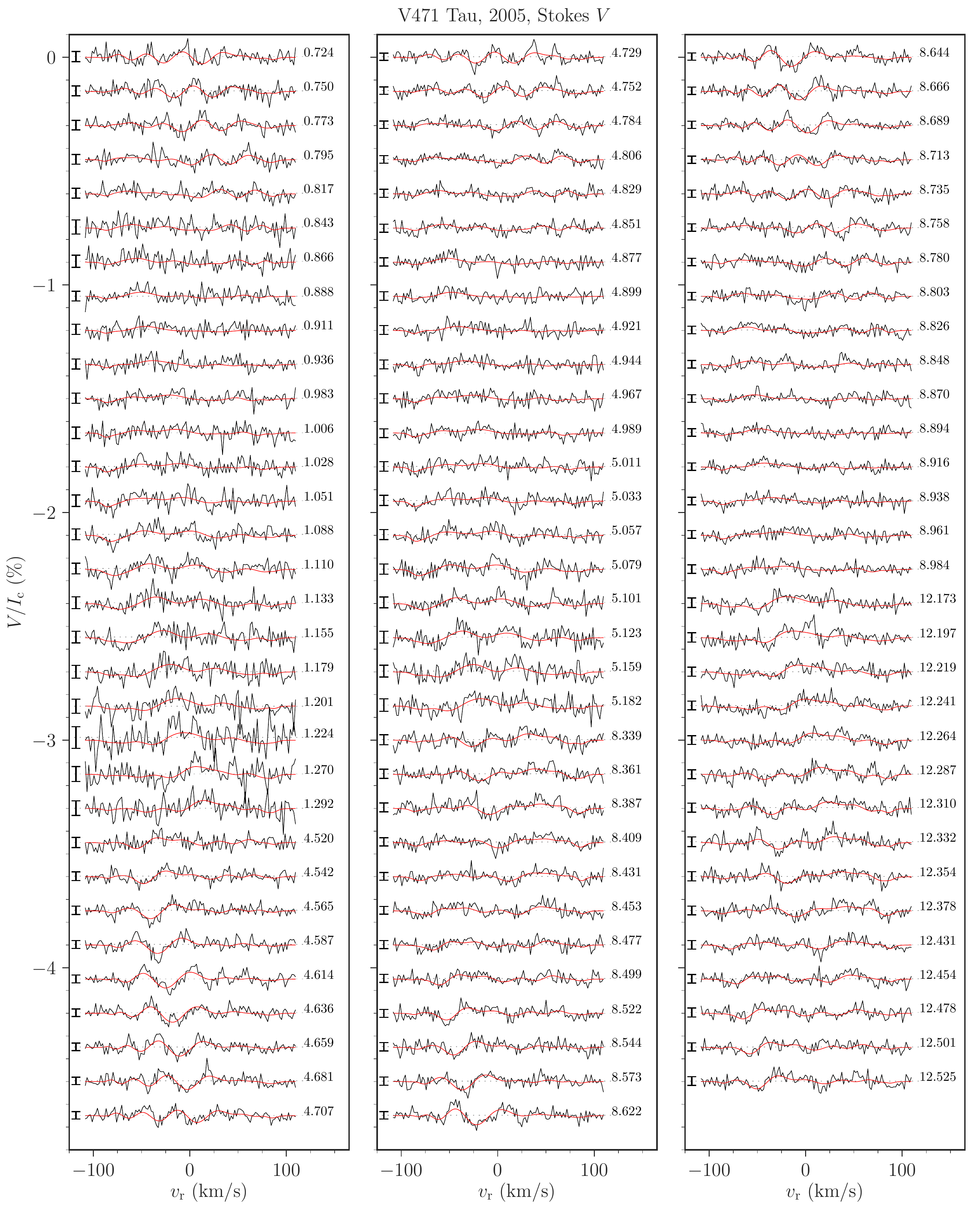}
    \caption{Same as Fig.~\ref{fig:sv2004} but for the $95$ 
    Stokes $V$ spectra collected in Dec 2005.}
    \label{fig:sv2005}
\end{figure*}

Figures~\ref{fig:si2004} and \ref{fig:si2005} show the variations of the unpolarized profiles for Nov/Dec 2004 and Dec 2005, respectively. The observed profiles are shown as black lines and the maximum-entropy fit as red lines. Similarly, circularly polarized profiles are shown in figs.~\ref{fig:sv2004} (Nov/Dec 2004) and \ref{fig:sv2005} (Dec 2005).

\section{Proxies of magnetic activity}\label{sec:lines} 
\begin{figure*}
    \centering
    \includegraphics[width=0.3\textwidth]{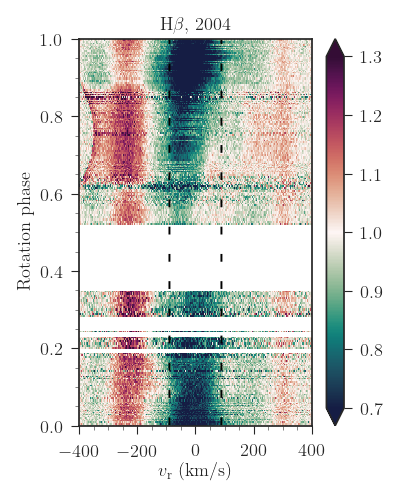}
    \includegraphics[width=0.3\textwidth]{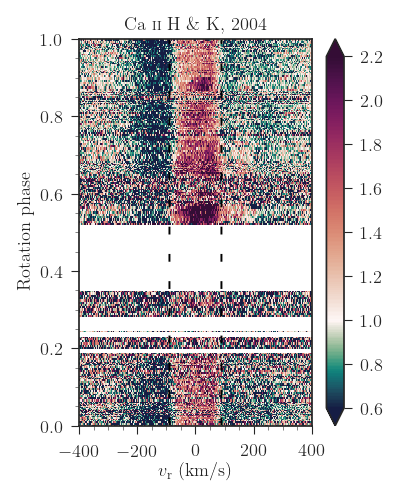} 
    \includegraphics[width=0.3\textwidth]{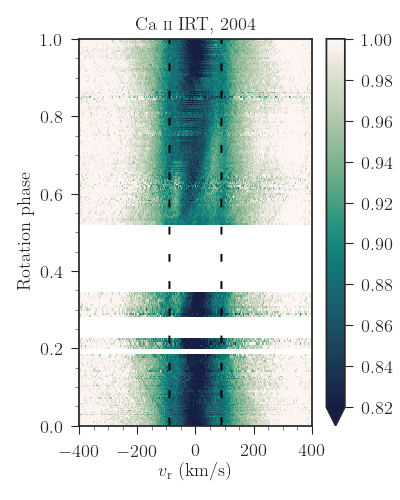}
    \includegraphics[width=0.3\textwidth]{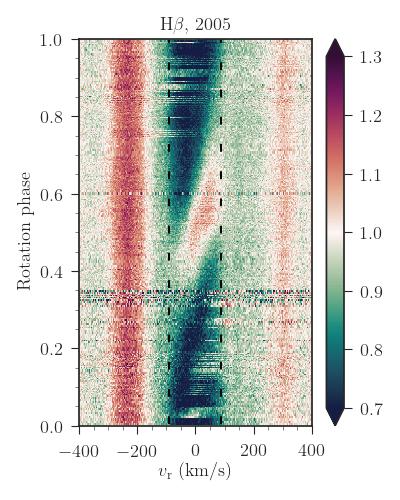}
    \includegraphics[width=0.3\textwidth]{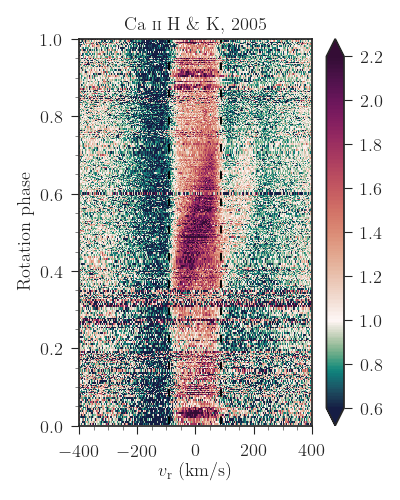}
    \includegraphics[width=0.3\textwidth]{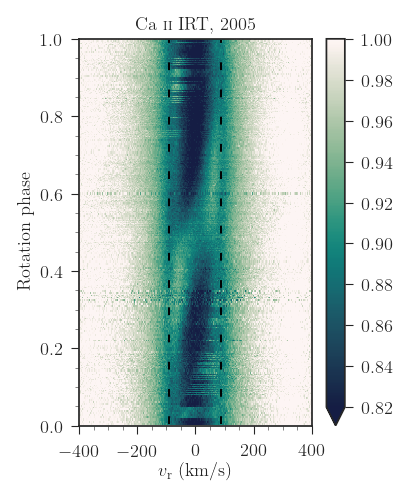}
    \caption{Dynamical spectra collected in Nov/Dec 2004 (top panel) and Dec 2005 (bottom panel). From left to right,  H$\beta$, \ion{Ca}{ii} H $\&$ K and \ion{Ca}{ii} IRT lines are shown in the rest frame of the K2 dwarf. The dashed vertical lines illustrate velocities of $\pm v\sin i$. Note that H$\beta$ is blended with two emission lines located at $-240$~km/s ($\sim\negmedspace485.7$~nm) and 300~km/s ($\sim\negmedspace486.7$~nm).}
    \label{fig:activitylines}
\end{figure*}

Figure~\ref{fig:activitylines} displays the activity indicators H$\beta$, \ion{ca}{ii} H$\&$K and \ion{ca}{ii} infrared triplet for both observing epochs. 

%%%%%%%%%%%%%%%%%%%%%%%%%%%%%%%%%%%%%%%%%%%%%%%%%%

% Don't change these lines
\bsp	% typesetting comment
\label{lastpage}
\end{document}